\documentclass[12pt]{article}  
\usepackage{a4,cite,graphics,placeins} 

\def\normalspace{\def\baselinestretch{1.0}\normalsize}
\def\Address#1#2{$^{\rm#1}${\it\footnotesize#2}\\}
\def\Ref#1{(\ref{#1})}
\def\PSfig#1{\scalebox{0.75}{\includegraphics{#1}}}
\def\Caption#1{
  \normalspace
  \vskip -4mm
  \begin{quotation}
  \caption{\sl #1}
  \end{quotation}
  \vskip -12mm
}
\def\Appendix{\par
  \def\thesection{Appendix \Alph{section}}
  \def\thesubsection{\Alph{section}.\arabic{subsection}}
  \def\theequation{\Alph{section}.\arabic{equation}}
  \setcounter{section}{0}
  \setcounter{subsection}{0}
  \setcounter{equation}{0}
}
 
\def\Ref#1{(\ref{#1})}
\def\BA{\begin{eqnarray}}
\def\BE{\begin{equation}}
\def\BF{\begin{figure}[htb]}
\def\BT{\begin{table}[htb]}
\def\EA{\end{eqnarray}}
\def\EE{\end{equation}}
\def\EF{\end{figure}}
\def\ET{\end{table}}
\def\bb{\mbox{\boldmath$b$\unboldmath}}
  \def\vrT{{\vec r_T}}

\def\la{\langle}
\def\ra{\rangle}

\def\GeV{\,\mbox{GeV}}
\def\MeV{\,\mbox{MeV}}

\def\Jpsi{{J\!/\!\psi}}
\def\sqq{{\sigma_{q\bar q}}}

\def\eff{{\mbox{\tiny eff}}}

\def\max{{\mbox{\tiny max}}}

\def\tx{{\tilde x}}
\def\tQ{{\tilde Q}}
\def\tb{{\tilde b}}

\begin{document}

\title{\bf Electroproduction of Charmonia off Nuclei}
\author{
  Yu.P.~Ivanov$^{a,b,c}$,
  B.Z.~Kopeliovich$^{b,c,d}$,
  A.V.Tarasov$^{b,c,d}$
  and J.~H\"ufner$^{a,b}$
  \\
  \\
  \Address{a}{
    Institut f\"ur Theoretische Physik der Universit\"at, 
    Philosophenweg 19, 69120 Heidelberg, Germany
  }
  \Address{b}{
    Max-Planck Institut f\"ur Kernphysik,
    Postfach 103980, 69029 Heidelberg, Germany
  }
  \Address{c}{
    Joint Institute for Nuclear Research,
    Dubna, 141980 Moscow Region, Russia
  }
  \Address{d}{
    Institut f\"ur Theoretische Physik der Universit\"at,
    93040 Regensburg, Germany   
  }
}
\date{\today}
\maketitle
%\doublespace

\begin{abstract}
In a recent publication we have calculated elastic charmonium production
in $ep$ collisions employing realistic charmonia wave functions and
dipole cross sections and have found good agreement with the data in a
wide range of $s$ and $Q^2$. Using the ingredients from those
calculations we calculate exclusive electroproduction of charmonia off
nuclei. Here new effects become important, (i) color filtering of the
$c\bar c$ pair on its trajectory through nuclear matter, (ii) dependence
on the finite lifetime of the $c\bar c$ fluctuation (coherence length)
and (iii) gluon shadowing in a nucleus compared to the one in a nucleon.
Total coherent and incoherent cross sections for C, Cu and Pb as
functions of $s$ and $Q^2$ are presented together with some differential
cross sections. The results can be tested with future electron-nucleus
colliders or in peripheral collisions of ultrarelativistic heavy ions.
\end{abstract}

\newpage

\section{Introduction}

In contrast to hadro-production of charmonia, where the elementary
production vertex is still debated, electro(photo)production of charmonia
seems well understood: the $c\bar c$ fluctuation of the incoming real or
virtual photon interacts with the target (proton or nucleus) via the
dipole cross section $\sqq$ and the result is projected on the wave
function of the observed hadron. In the $\gamma^*p \to \Psi p$ reaction
only the dipole cross section $\sqq$ and the wave function of the $\Psi$
enter ($\Psi$ stands for $\Jpsi$ or $\psi'$). In comparing calculations
with experiment (which fortunately are rather precise and cover a wide
range of values $s$ and $Q^2$ of the incoming $\gamma^*$), various
parameterizations of the dipole cross section and the wave function have
been successfully tested \cite{HIKT}. In the exclusive electroproduction
of charmonia $\gamma^*A \to \Psi X$, where $X=A$ (coherent) or $X=A^*$
(incoherent, where $A^*$ is an exited state of the $A$-nucleon system)
new phenomena are to be expected: color filtering, i.e. inelastic
interactions of the $c\bar c$ pair on its way through the nucleus is
expected to lead to a suppression of $\Psi$ production relative to $A
\sigma_{\gamma^*p\to\Psi p}$. Since the dipole cross section $\sqq$ also
depends on the gluon distribution in the target ($p$ of $A$), nuclear
shadowing of the gluon distribution is expected to reduce $\sqq$ in a
nuclear reaction relative to the one on the proton. Production of a
$c\bar c$ pair in a nucleus and its absorption are also determined by the
values of the coherence length $l_c$ and the formation length $l_f$.

Explicit calculations have already been performed in \cite{KZ91} in the
approximation of a short coherence (or production) length, when one can
treat the creation of the colorless $c\bar c$ pair as instantaneous,
\BE
  l_c = \frac{2\,\nu}{M_{c\bar c}^2} \approx 
        \frac{2\,\nu}{M_{\Jpsi}^2}\ \ll R_A,
  \label{10}
\EE
where $\nu$ is the energy of the virtual photon in the rest frame of the
nucleus. At the same time, the formation length may be long, comparable
with the nuclear radius $R_A$,
\BE
  l_f = \frac{2\,\nu}{M_{\psi'}^2 - M_{\Jpsi}^2} \sim R_A\ .
  \label{20}
\EE
In \cite{KZ91} the wave function formation is described by means of the
light-cone Green function approach summing up all possible paths of the
$c\bar c$ in the nucleus. The result has been unexpected. Contrary to
naive expectation, based on the larger size of the $\psi'$ compared to
$\Jpsi$, it has been found that $\psi'$ is not more strongly absorbed
than the $\Jpsi$, but may even be enhanced by the nuclear medium. This
is interpreted as an effect of filtering which is easy to understand
in the limit of long coherence length, $l_f\gg R_A$, when the ratio of
cross sections on nuclear and nucleon targets take the simple form.
Indeed, the production rate of $\psi'$ on a proton target is small due
to strong cancellations in the projection of the produced $c\bar c$ wave
packet onto the radial wave function of the $\psi'$ which has a node.
After propagation through nuclear matter the transverse size of a
$c\bar c$ wave packet is squeezed by absorption and the projection of
the $\psi'$ wave function is enhanced\cite{KZ91,BKMNZ} since the effect
of the node is reduced.

However, the quantitative predictions of \cite{KZ91} are not trustable
since the calculations have been oversimplified and quite some progress
has been made on the form of the dipole cross section $\sqq$ and the 
light cone wave functions for the charmonia. Therefore we take the 
problem up again and provide more realistic calculations for nuclear
effects in exclusive electroproduction of charmonia off nuclei relying on the
successful parameter free calculations which have been performed recently
by the authors \cite{HIKT} for elastic virtual photoproduction of
charmonia, $\gamma^*\,p \to \Psi\,p$. Since the present paper is
a direct follow up of \cite{HIKT} we do not repeat all definitions
and notations, but rather request the reader to consult \cite{HIKT}
for more details.

Whenever one deals with high-energy reactions on nuclei, one cannot avoid
another problem of great importance: gluon shadowing. At small values of
$x$, gluon clouds overlap in longitudinal direction and may fuse. As a
result, the gluon density per one nucleon in a nucleus is expected to be
reduced compared to a free proton. Parton shadowing, which leads to an
additional nuclear suppression in various hard reactions (DIS, DY, heavy
flavor, high-$p_T$ hadrons, etc.) may be especially strong for exclusive
vector meson production like charmonium production which needs at least two
gluon exchange. Unfortunately, we have no experimental information for gluon
shadowing in nuclei so far, and we have to rely on the available theoretical
estimates \cite{Mueller90,Mueller99,KST,KRTJ,BBFS}.

In the present paper we work in the approximation of long coherence time
$l_c\gg R_A$ since calculations can be done with great confidence. Only
recently \cite{KNST} the light-cone (LC) dipole approach has been
generalized also for the case of a finite coherence length, relevant for
the vast majority of available data. However, because of technical
difficulties those calculations employ the $\propto r_T^2$ shape of the
dipole cross section. It is still a challenge to perform calculations
with a realistic dipole cross section for the regime of finite coherence
length. Nevertheless, we make corrections for finite values of $l_c$
employing the approximation suggested in \cite{BKMNZ}.

The predictions in this paper are meant as a realistic basis for the
planning of future experiments for electron-nucleus collisions at high
energies like in the eRHIC project. It would be highly desirable to have
good data in order to test the assumptions and ingredients which have
entered the calculation and which are important aspects of modern strong
interaction many-body physics. Furthermore, progress in the understanding
of charmonium electroproduction will also have important consequences for
hadroproduction.

The paper is organized as follows. Section~\ref{section-nuclei}
introduces some notations and the most important definitions and
expressions. Then the integrated cross sections for coherent and
incoherent charmonia production are calculated as function of $s$ and
$Q^2$ of the virtual photon and the results are presented as a ratio to
$A$-times the cross sections on a proton. In the following two sections,
two important modifications are introduced. The effect of gluon shadowing
in the dipole cross section (Section~\ref{section-gluon}) and the case of
finite values of $l_c$ ($l_c \simeq R_A$) (Section~\ref{section-lc}).
Appendix A describes some very useful calculational procedures, which
considerably increase the accuracy and the speed of numerical
calculations.

\section{
  Shadowing and absorption for \boldmath$c\bar c$ pairs in nuclei
  \label{section-nuclei}
}

Exclusive charmonium production off nuclei, $\gamma^* A \to \Psi X$ is
called coherent, when the nucleus remains intact, i.e. $X=A$, or
incoherent, when $X$ is an excited nuclear state which contains nucleons
and nuclear fragments but no other hadrons. The cross sections depend on
the polarization $\epsilon$ of the virtual photon (in all figures below
we will imply $\epsilon=1$),
\BE
  \sigma^{\gamma^*A}(s,Q^2) =
  \sigma^{\gamma^*_TA}(s,Q^2) + \epsilon\,
  \sigma^{\gamma^*_LA}(s,Q^2)~,
  \label{30}
\EE
where the indexes $T,L$ correspond to transversely or longitudinally 
polarized photons, respectively.

The cross section for exclusive production of charmonia off a nucleon 
target integrated over momentum transfer \cite{ZKL} is given by
\BE
  \sigma^{\gamma_{T,L}^*N}_{inc}(s,Q^2) =
  \left|\left\la\Psi\left|\sqq(r_T,s)
  \right|\gamma^{T,L}_{c\bar c}
  \right\ra\right|^2~,
  \label{35} 
\EE 
where $\Psi(\vec r_T,\alpha)$ is the charmonium LC wave function which
depends on the transverse $c\bar c$ separation $\vec r_T$ and on the
relative sharing $\alpha$ of longitudinal momentum \cite{HIKT}. Both
variables are involved in the integration in the matrix element
Eq.~\Ref{35}. $\Psi(\vec r_T,\alpha)$ is obtained by means of a Lorentz
boost applied the solutions of the Schr\"odinger equation. This procedure
involves the Melosh spin rotation \cite{Terent'ev,Melosh} which produces
sizable effects.

In Eq.~\Ref{35} $\gamma^{T,L}_{c\bar c}(\vec r_T,\alpha,Q^2)$ is the
LC wave function of the $c\bar c$ Fock component of the photon. It
depends on the photon virtuality $Q^2$.  One can find the details in
Ref.~\cite{HIKT} including the effects of a nonperturbative $q\bar q$
interaction.

The cross section $\sqq(r_T,s)$ describes the interaction of a colorless
quark-antiquark dipole of separation $r_T$ and c.m. energy squared $s$
with a nucleon. At small values of $r_T$ the energy dependence should
come in the combination $x\sim 1/(r_T^2s)$ in order to respect Bjorken
scaling. A simple parameterization of $\sqq(r_T,x)$ suggested in
\cite{GBW} (GBW) well describes the data for $F_2^p(x,Q^2)$ at small $x$
and high $Q^2$. At low $Q^2$, however, the energy variable $s$ is more
appropriate. A parameterization for $\sqq(r_T,s)$ suggested in \cite{KST}
(KST) describes DIS data only up to $Q^2\sim 10\,\GeV^2$ which is a scale
relevant for charmonia. Since we cannot see why one parameterization should
be better than the other, we calculate results for both parameterizations as
has been done in \cite{HIKT}.

The cross sections for coherent and incoherent production on nuclei will
be derived under various conditions imposed by the coherence length
Eq.~(\ref{10}). At high energies the coherence length Eq.~\Ref{10} may
substantially exceed the nuclear radius. In this case the transverse size
of the $c\bar c$ wave packet is ``frozen'' by Lorentz time dilation, i.e.
it does not fluctuate during propagation through the nucleus, and the
expressions for the cross sections, incoherent ($inc$) or coherent
($coh$), are particularly simple \cite{KZ91},
\BA
  \sigma^{\gamma_{T,L}^*A}_{inc}(s,Q^2) &=& 
  \int d^2b\,T_A(b)\,
  \left|\left\la\Psi\left|\sqq(r_T,s)\,
  \exp\left[ -{1\over2}\, \sqq(r_T,s)\,T_A(b)\right]
  \right|\gamma^{T,L}_{c\bar c}
  \right\ra\right|^2
  \label{40}\\
  \sigma^{\gamma_{T,L}^*A}_{coh}(s,Q^2) &=&
  \int d^2b\,\left|\left\la\Psi\left|1\,-\,
  \exp\left[-{1\over2}\,\sqq(r_T,s)\,T_A(b)\right]
  \right|\gamma^{T,L}_{c\bar c}
  \right\ra\right|^2\ .
  \label{50}
\EA
Here $T_A(b)=\int_{-\infty}^{\infty}dz\,\rho_A(b,z)$ is the nuclear
thickness function given by the integral of the nuclear density along the
trajectory at a given impact parameter $b$. 

In Eqs.~\Ref{40} and \Ref{50} we also include a small correction due to
the real part of the $\gamma^*N\to\Psi N$ amplitude via replacement
\BE
  \sqq\Rightarrow 
  \sqq\,\left(1-\frac{i\pi}2\frac{\partial\ln\sqq}{\partial\ln s}\right)~.
  \label{60}
\EE

As it is common practice we express nuclear cross sections for
incoherent and coherent production in the form of the ratio,
\BE
  R_\Psi(s,Q^2) =
  \frac{\sigma^{\gamma^*A}(s,Q^2)}{A\,\sigma^{\gamma^*N}(s,Q^2)}~,
  \label{70}
\EE
where the numerator stands for the expressions Eqs.~\Ref{40} and \Ref{50}
for incoherent and coherent cross sections, respectively. The details of
the technically rather complicated numerical calculations of the integrals
in Eqs.~\Ref{40} and \Ref{50} can be found in \ref{section-resumm}.

The ratio Eq.~\Ref{70} for incoherent electroproduction of $\Jpsi$ and
$\psi'$ is shown in Fig.~\ref{s-inc} as a function of $\sqrt{s}$. We use
the GBW \cite{GBW} and KST \cite{KST} parameterizations for the dipole
cross section and show the results by solid and dashed curves,
respectively. Differences are at most $10-20\,\%$.
\BF
\centerline{
  \PSfig{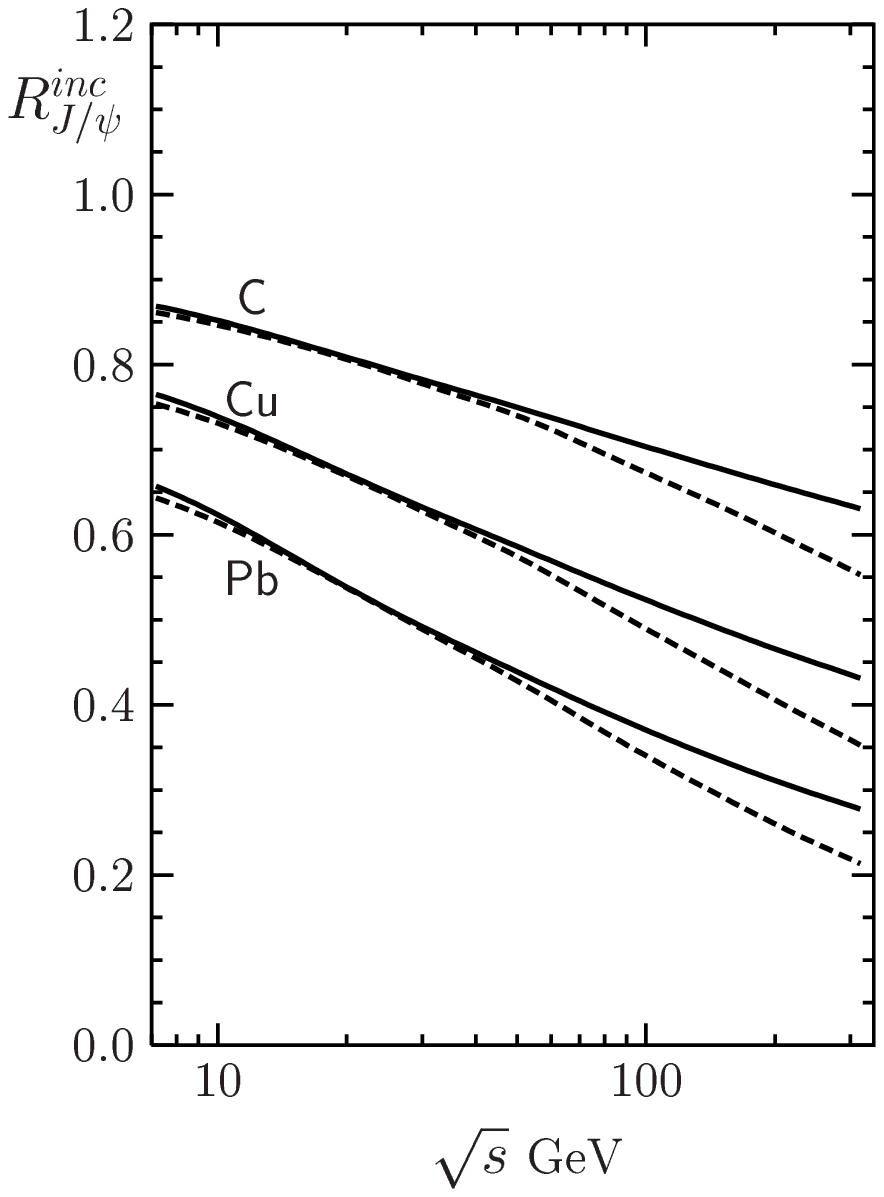}\hskip5mm
  \PSfig{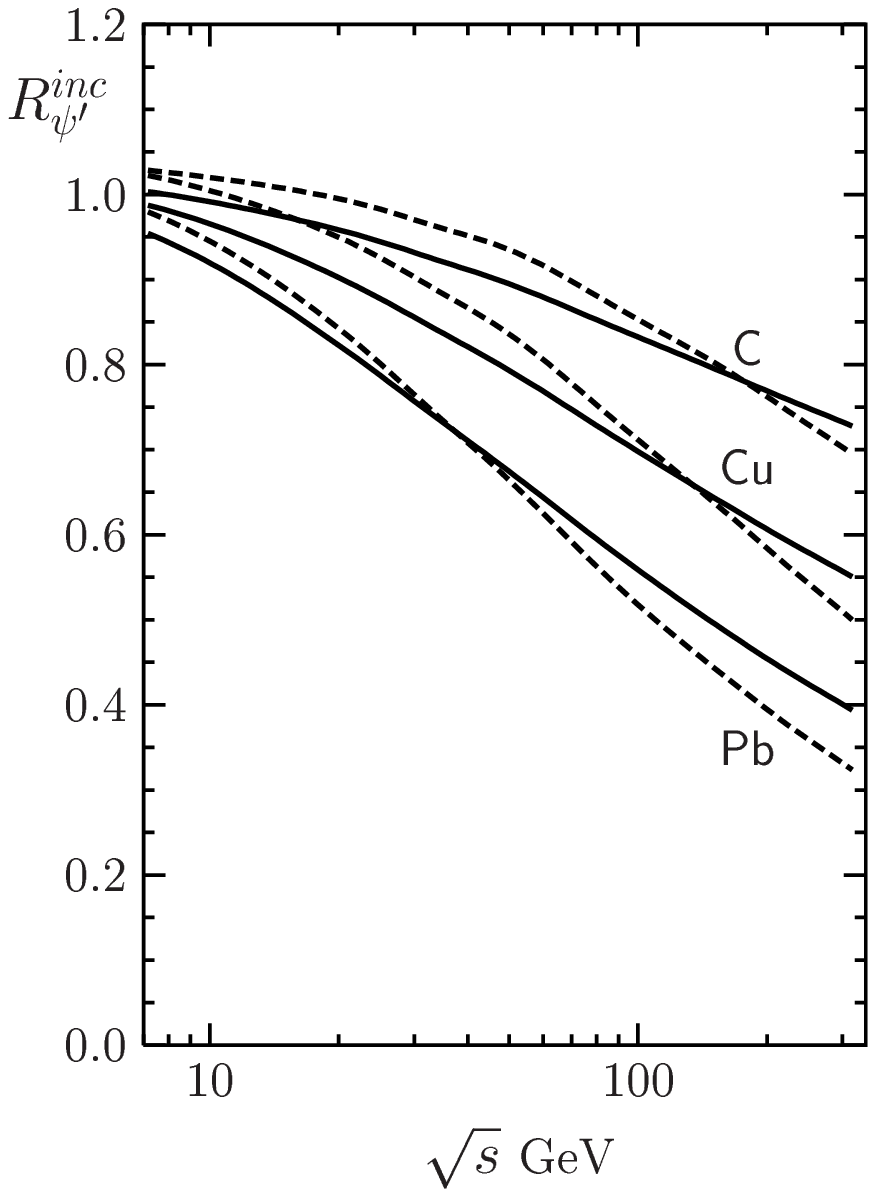}
}
\Caption{
  \label{s-inc}
  Ratios $R^{inc}_\Psi$ for $\Jpsi$ and $\psi'$ incoherent production
  on carbon, copper and lead as function of $\sqrt s$ and at $Q^2=0$.
  The solid curves refer to the GBW parameterization of $\sqq$ and
  dashed one refer to the KST parameterization.
}
\EF

Analyzing the results shown in Fig.~\ref{s-inc}, we observe that nuclear
suppression of $\Jpsi$ production becomes stronger with energy. This is
an obvious consequence of the energy dependence of $\sqq(r_T,s)$, which
rises with energy (see in Ref.~\cite{HIKT}). For $\psi'$ the suppression
is rather similar as for $\Jpsi$. In particular we do not see any
considerable nuclear enhancement of $\psi'$ which has been found earlier
\cite{KZ91,KNNZ}, where the oversimplified form of the dipole cross
section, $\sqq(r_T)\propto r_T^2$ and the oscillator form of the wave
function had been used. Such a form of the cross section enhances the
compensation between large and small distances in the wave function of
$\psi'$ in the process $\gamma^*p\to\psi' p$. Therefore, the color
filtering effect which emphasizes the small distance part of the wave
function leads to a strong enhancement of the $\psi'$ production rate.
This is why using the more realistic $r_T$-dependence of $\sqq(r_T)$
leveling off at large $r_T$ leads to a weaker enhancement of the $\psi'$.
This effect becomes even more pronounced at higher energies since the 
dipole cross section saturates starting at a value $r_T \sim r_0(s)$
where $r_0(s)$ decreases with energy. This observation probably explains
why the $\psi'$ is less enhanced at higher energies as one can see from
Fig.~\ref{s-inc}.

Note that the ``frozen'' approximation is valid only for $l_c\gg R_A$ and
can be used only at $\sqrt{s} > 20-30\,\GeV$. Therefore, the low-energy
part of the curves depicted in Fig.~\ref{s-inc} should be corrected for
the effects related to the finiteness of $l_c$. This will be done in
Section~\ref{section-lc}.

One can change the effect of color filtering in nuclei in a controlled way
by increasing the photon virtuality $Q^2$ thereby squeezing the transverse
size of the $c\bar c$ fluctuation in the photon. For a narrower $c\bar c$ 
pair the cancellation which is caused by the node in the radial wave function
of $\psi'$ should be less effective. One expects that the $\psi'$ to $\Jpsi$
ratio on a proton target increases with $Q^2$, as is observed both in
experiment and calculation (Fig.~9 of \cite{HIKT}). Fig.~\ref{Q-inc}
shows the result for nuclear targets: for large values of $Q^2$ ratios
$R^{inc}_\Jpsi$ and $R^{inc}_{\psi'}$ become very similar. This effect
has also been observed in the E665 experiment \cite{E665} for exclusive
production of $\rho$ and $\rho'$ off nuclei at high energies.
\BF
\centerline{
  \PSfig{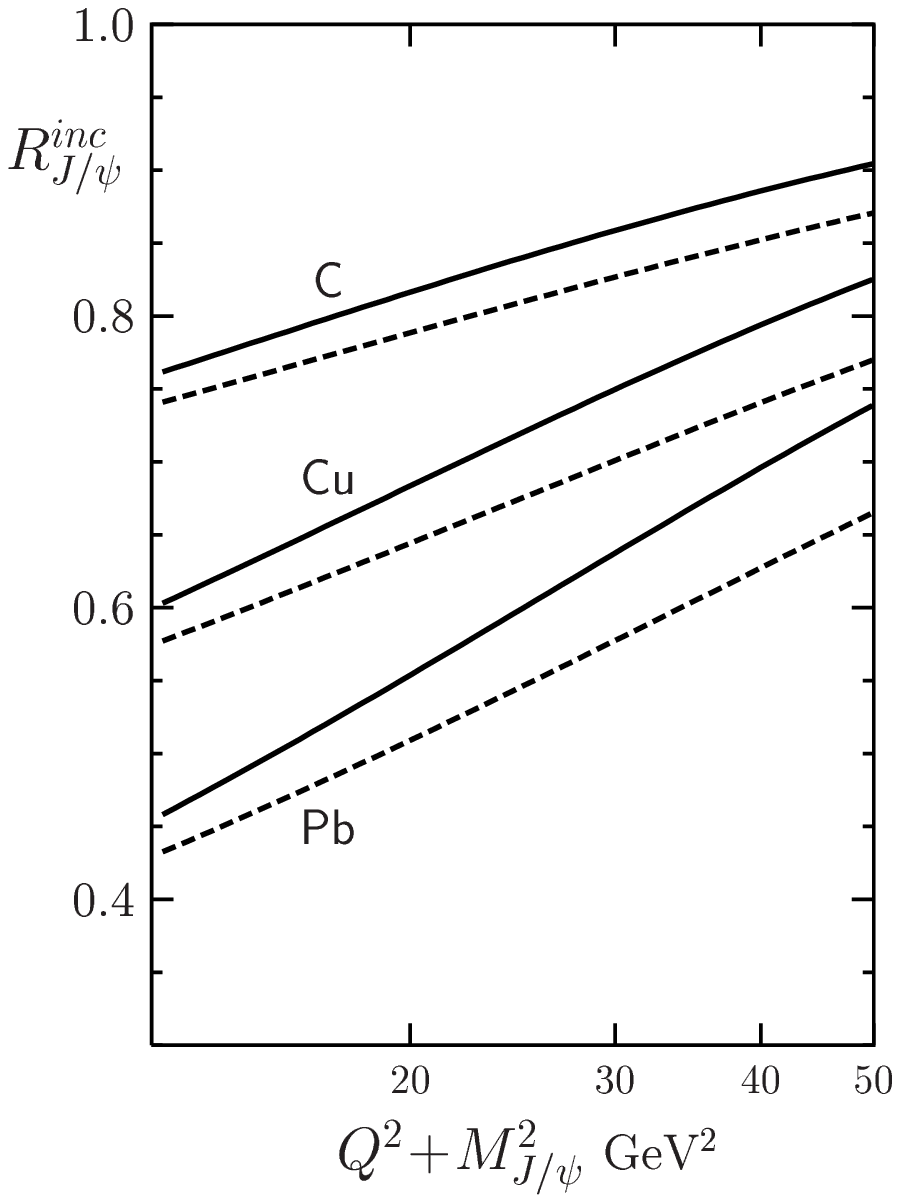}\hskip5mm
  \PSfig{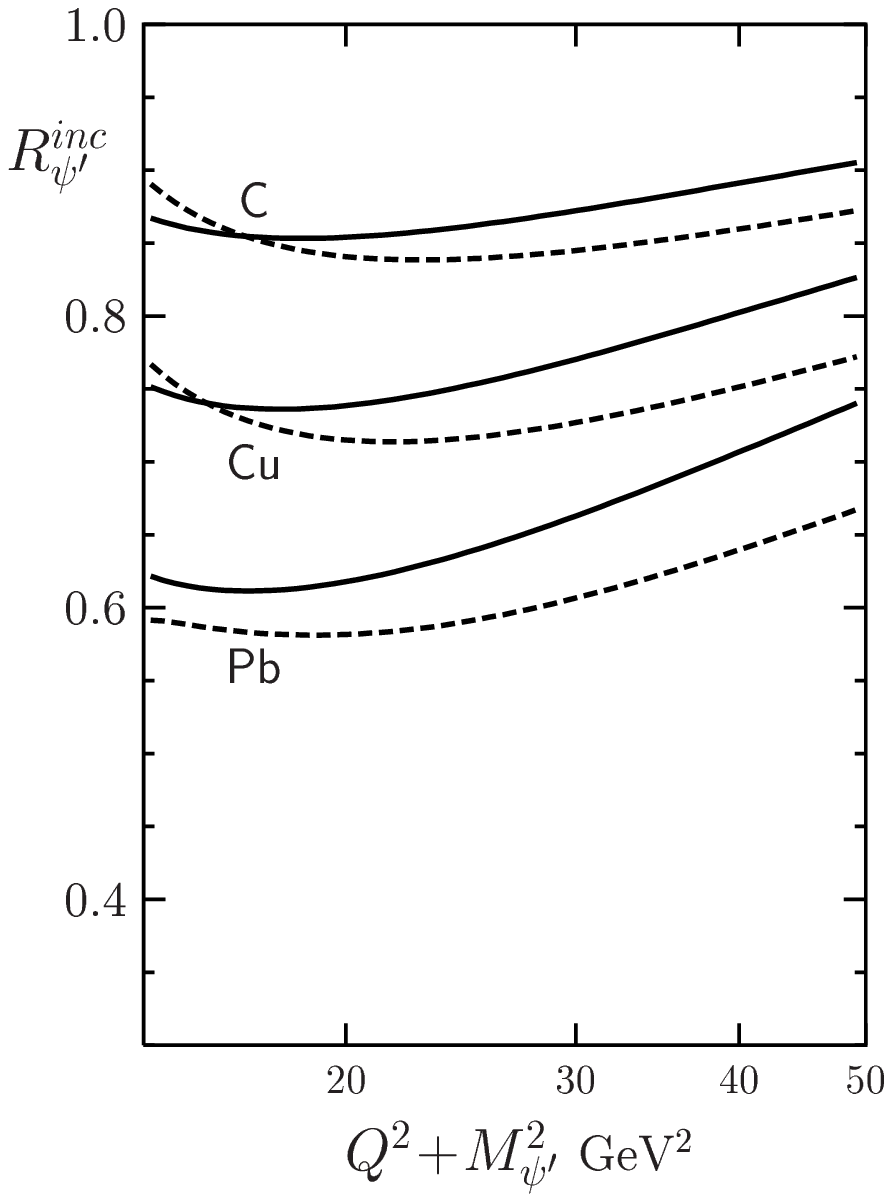}
}
\Caption{
  \label{Q-inc}
  Ratios $R^{inc}_{\Jpsi}$ and $R^{inc}_{\psi'}$ for incoherent
  production on three nuclei displayed as a function of $Q^2+M_\Psi^2$
  at fixed $s=4000\GeV^2$. The lines are for the GBW (solid curves)
  and KST (dashed curves) parameterizations.
}
\EF

Nuclear effects for $\psi'$ production shown in Fig.~\ref{Q-inc}
demonstrate an even more peculiar $Q^2$ dependence. The overlap of the
produced $c\bar c$ state and the $\psi'$ wave function rises with $Q^2$
both in the numerator and denominator of the ratio Eq.~\Ref{70} due to
the node effect. However, the nuclear filtering for large size $c\bar c$
pairs is especially strong at small $Q^2$, hence the $Q^2$ squeezing does
not affect the numerator as much as the denominator. This is why the
ratio $R_{\psi'}^{inc}(Q^2)$ can be a falling function at small $Q^2$. At
high $Q^2$, however, the size of the $c\bar c$ is so small that color
filtering is not an important effect any more, while the nucleus becomes
more transparent eventually causing $R_{\psi'}^{inc}(Q^2)$ to rise.

Cross sections for coherent production of charmonia on nuclei are calculated
analogously using Eq.~\Ref{50}. The results for the energy dependence are
depicted in Fig.~\ref{s-coh} and for the $Q^2$ dependence in Fig.~\ref{Q-coh}.
\BF
\centerline{
  \PSfig{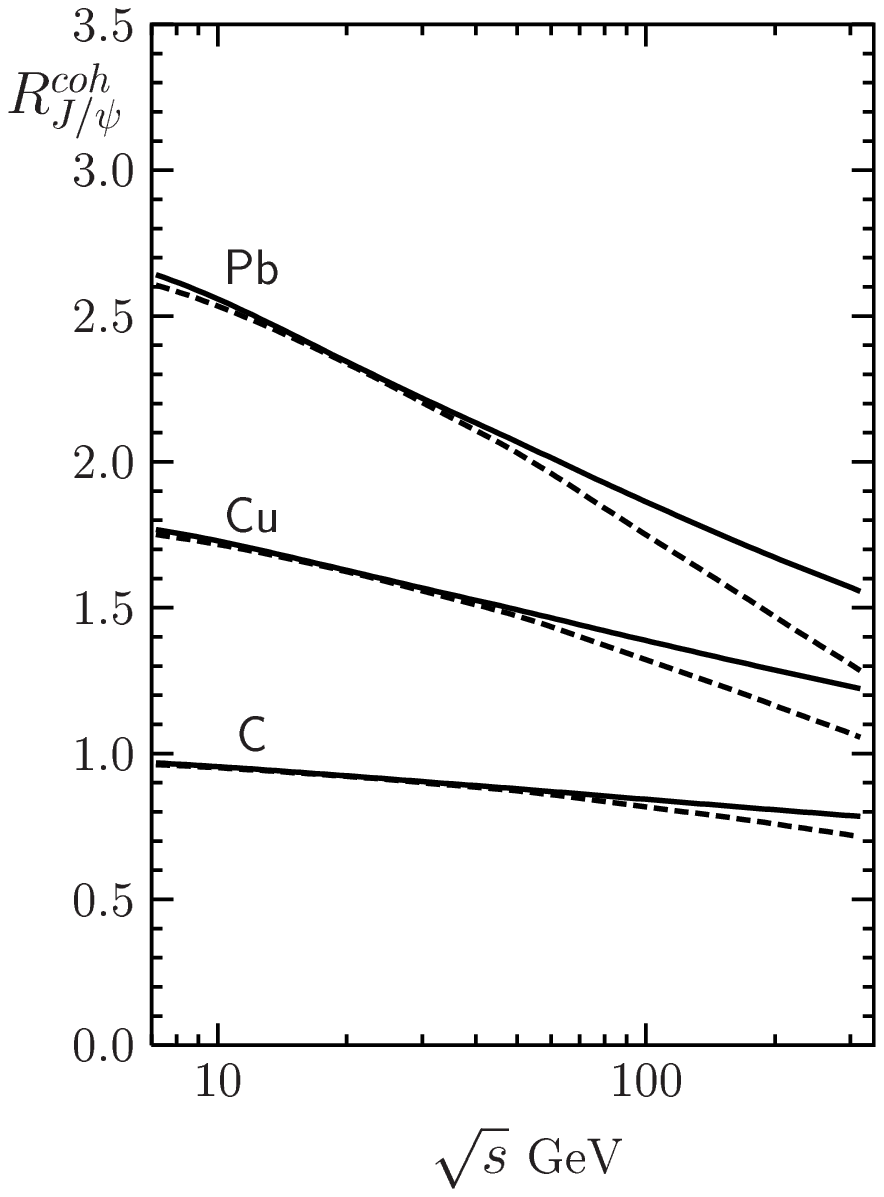}\hskip5mm
  \PSfig{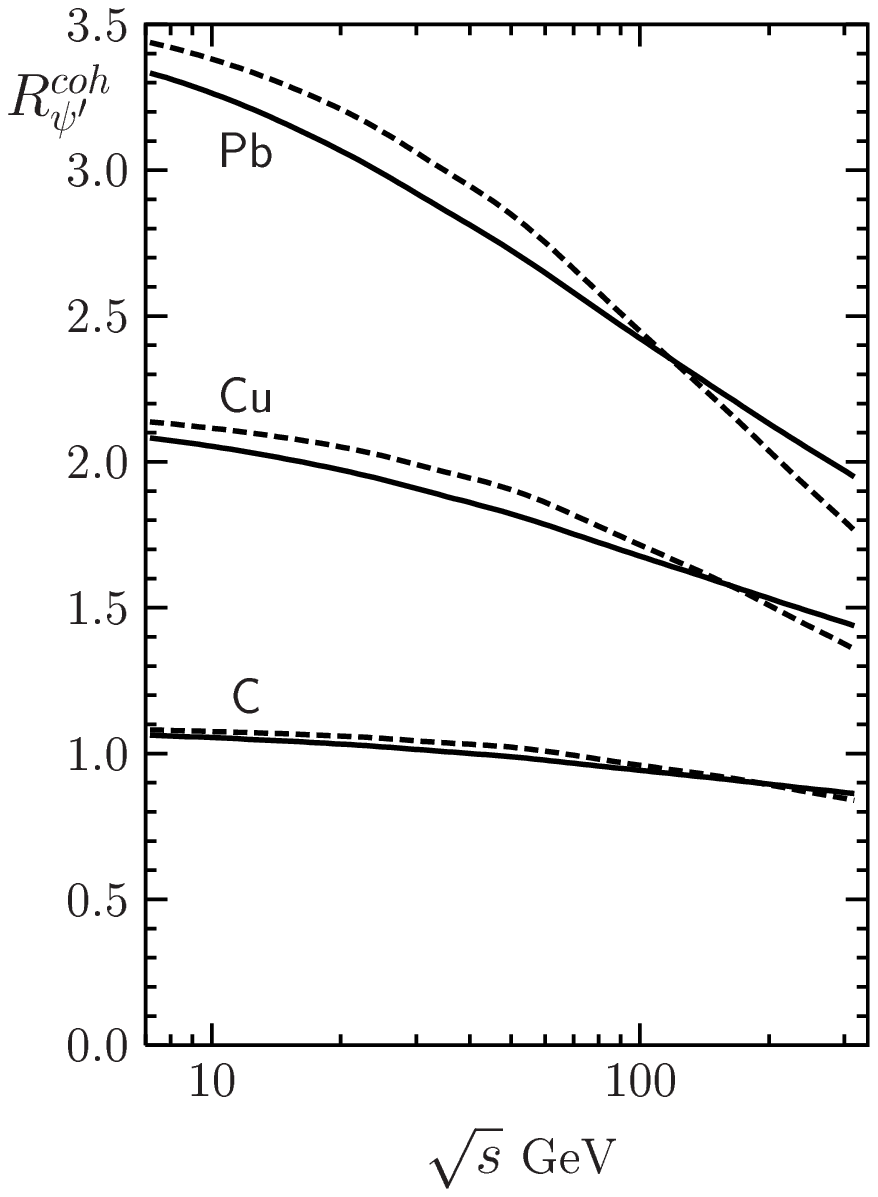}
}
\Caption{
  \label{s-coh}
  The ratios $R^{coh}_\Jpsi$ and $R^{coh}_{\psi'}$ for coherent
  production on nuclei as a function of $\sqrt s$. The meaning of
  the different lines is the same as in Fig.~\ref{s-inc}.
}
\EF
\BF
\centerline{
  \PSfig{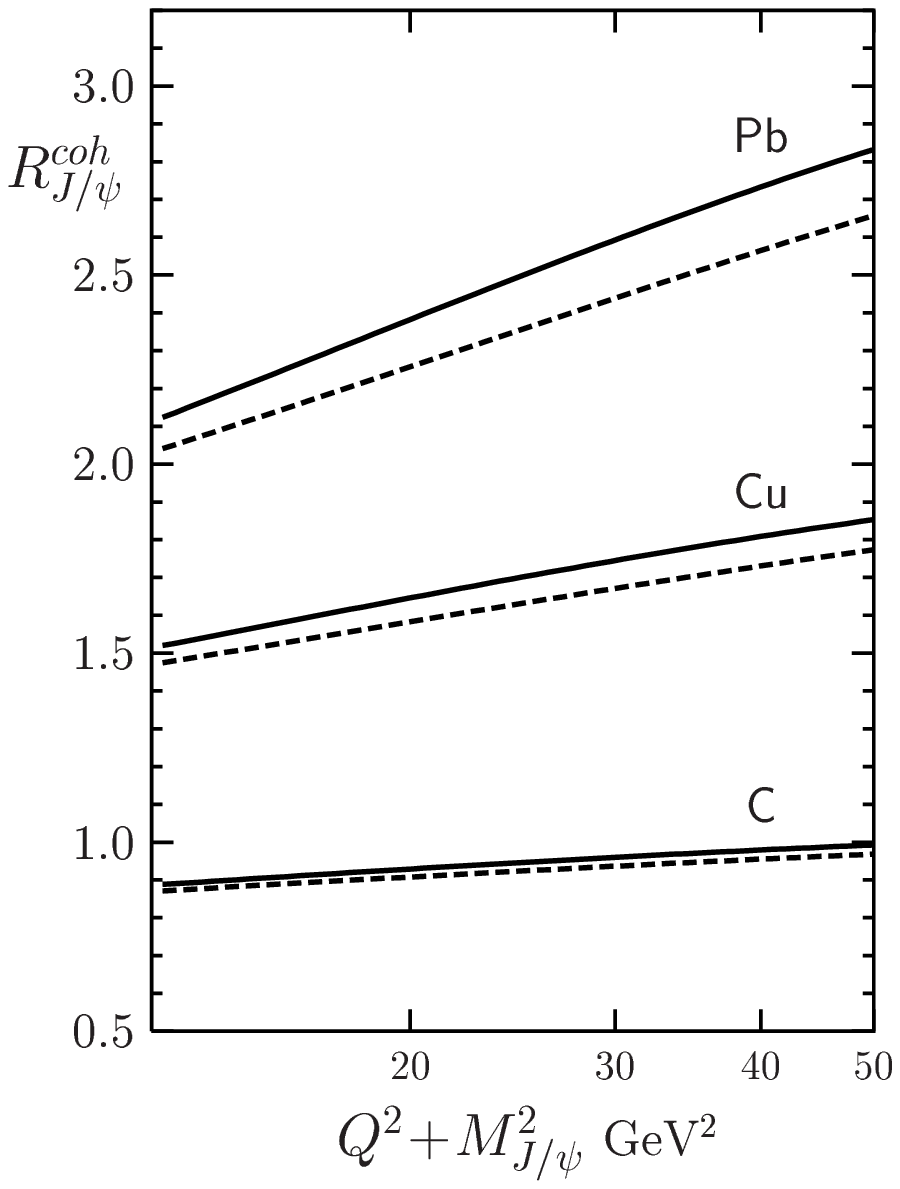}\hskip5mm
  \PSfig{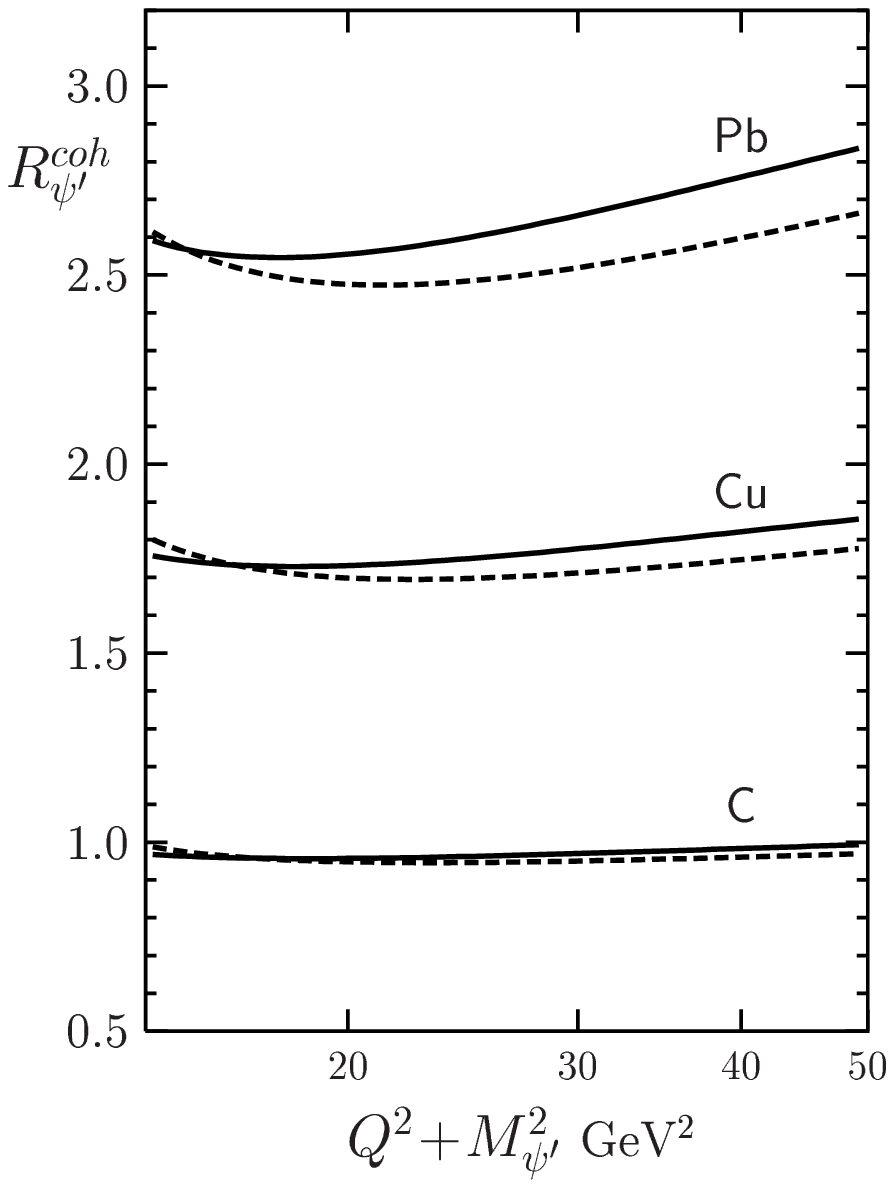}
}
\Caption{
  \label{Q-coh}
  The ratios $R^{coh}_\Jpsi$ and $R^{coh}_{\psi'}$ for coherent
  production on various nuclei as a function of $Q^2$. The meaning
  of the lines is the same as in Fig.~\ref{Q-inc}.
}
\EF
It is not a surprise that the ratios exceed one. In the absence of
$c\bar c$ attenuation the forward coherent production would be
proportional to $A^2$, while integrated over momentum transfer,
the one depicted in Fig.~\ref{s-coh}, behaves as $A^{4/3}$. It is
a result of our definition Eq.~\Ref{70} that $R^{coh}_{\Psi}$
exceeds one. To avoid this artifact one may redefine the ratio
for coherent processes as,
\BE
  \widetilde R^{coh}_{\Psi}(s,Q^2)
  = \frac{\sigma_{coh}^{\gamma^*A}(s,Q^2)}
    {A\,\la T_A\ra \,16 \pi B^{\gamma^*N}\sigma^{\gamma^* N}(s,Q^2)}
  = \frac{R^{coh}_{\Psi}(s,Q^2)}{\la T_A\ra \,16 \pi B^{\gamma^*N}}~,
  \label{80}
\EE
where $B^{\gamma^*N}$ is the slope of the $\gamma^*N\to\Psi N$ differential
cross section and
\BE
  \la T_A \ra = {1\over A}\,\int\!d^2b\,T^2_A(b)
  \label{85}
\EE
is the mean nuclear thickness. This ratio approaches one when nuclear
effects disappear. Nevertheless, we will follow the usual definition
Eq.~\Ref{70} which is widely used for the presentation of data.

We can also predict the dependence on the momentum transfer $\vec k_T$
for the charmonium electroproduction on nuclei. In the case of incoherent
production this dependence is the same as for production on free nucleons.
However, in coherent production the nuclear formfactor comes into play
and one has
\BE
  \frac{d\sigma^{\gamma_{T,L}^*A}_{coh}(s,Q^2)}{d^2k_T} =
  \left| \int\!\!d^2b \,\, e^{i\vec k_T\cdot\vec b}\,
  \left\la \Psi\left|1\,-\,\exp\left[-{1\over2}\,
  \sqq(r_T,s)\,T_A(b)\right]\right|\gamma^{T,L}_{c\bar c}
  \right\ra\right|^2\ .
  \label{90}
\EE
In Fig.~\ref{kT-dep} we plot the ratios of total distributions (sum
of $T$ and $L$ components Eq.~\Ref{90} in the form of Eq.~\Ref{30}
normalized at $Q^2=0$ and $k_T=0$),
\BE
  {\cal R}(s,Q^2,k_T) =
  \frac{d\sigma^{\gamma^*A}_{coh}(s,Q^2)}{d^2k_T} \left/\left.
  \frac{d\sigma^{\gamma^*A}_{coh}(s,Q^2=0)}{d^2k_T}
  \right\vert_{k_T=0}\right.
\EE
as functions of $k_T$ at $s=4000\GeV^2$ for different virtualities
$Q$ of the photon.
\BF
\centerline{
  \PSfig{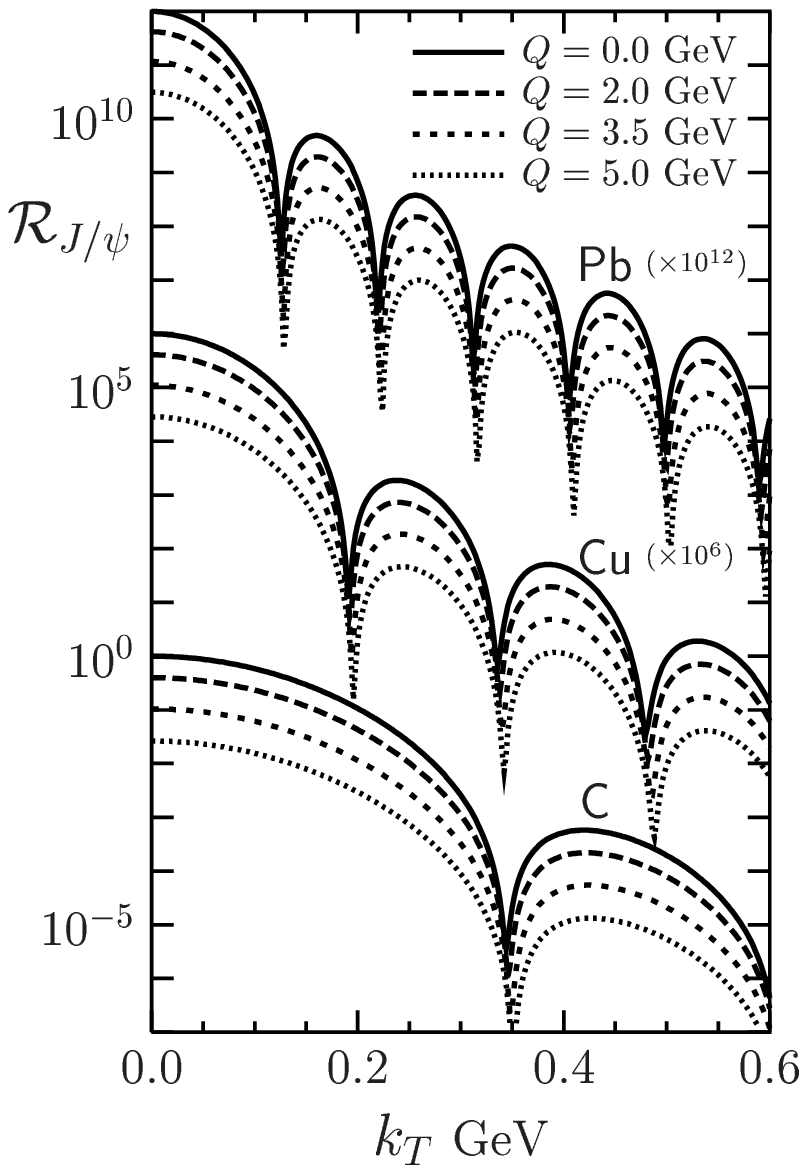}\hskip5mm
  \PSfig{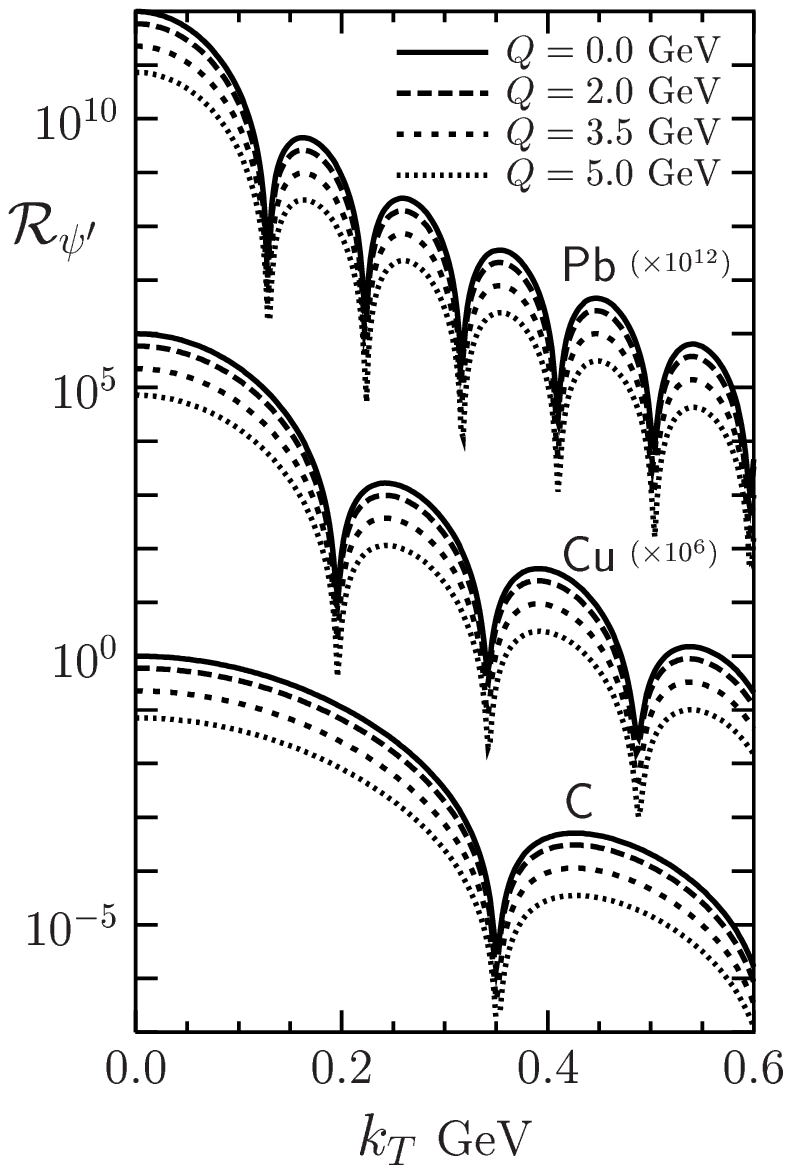}
}
\Caption{
  \label{kT-dep} Ratios ${\cal R}_\Jpsi$ and ${\cal R}_{\psi'}$ as
  functions of $k_T$ at $s=4000\GeV^2$ for different values of $Q$.
  All curves are calculated with the GBW parameterization of the dipole
  cross section $\sqq$.
}
\EF
We see that the $k_T$ dependences are rather similar for $\Jpsi$ and
$\psi'$. The shape of the distribution is governed mainly by the nuclear
geometry (and not by the size of the (small) charmonium). The calculated
curves show the familiar diffraction pattern known from elastic scattering
on nuclei.

\section{Gluon shadowing\label{section-gluon}}

The gluon density in nuclei at small Bjorken $x$ is expected to be
suppressed compared to a free nucleon due to interferences. This
phenomenon called gluon shadowing renormalizes the dipole cross section,
\BE
  \sqq(r_T,x) \Rightarrow \sqq(r_T,x)\,R_G(x,Q^2,b)\,.
  \label{glue}
\EE
where the factor $R_G(x,Q^2,b)$ is the ratio of the gluon density at $x$
and $Q^2$ in a nucleon of a nucleus to the gluon density in a free nucleon.
No data are available so far which could provide direct information about 
gluon shadowing. Currently it can be evaluated only theoretically.
In what follows we employ the technique developed in \cite{KST}.

The interpretation of the phenomenon of gluon shadowing depends very
much on the reference frame. It looks like glue-glue fusion in the
infinite momentum frame of the nucleus: although the nucleus is Lorentz
contracted, the bound nucleons are still well separated since they
contract too. However, the gluon clouds of the nucleons are contracted
less since they have a smaller momentum fraction $\sim x$. Therefore,
they do overlap and interact at small $x$, and gluons originating from
different nucleons can fuse leading to a reduction of the gluon density.

Although observables must be Lorentz invariant, the space-time
interpretation of shadowing looks very different in the rest frame of
the nucleus. Here it comes as a result of eikonalization of higher Fock
components of the incident particles. Indeed, the nuclear effect included
via eikonalization into Eqs.~(\ref{40})-(\ref{50}) corresponds to the
lowest $c\bar c$ Fock component of the photon. These expressions do not
include any correction for gluon shadowing, but rather correspond to
shadowing of sea quarks in nuclei, analogous to what is measured in
deep-inelastic scattering. Although the phenomenological dipole cross
section $\sqq(x,Q^2)$ includes all possible effects of gluon radiation,
the eikonal expressions Eqs.~(\ref{40})-(\ref{50}) assume that none of
the radiated gluons takes part in multiple interaction in the nucleus.
The leading order correction corresponding to gluon shadowing comes from
eikonalization of the next Fock component which contains the $c\bar c$
pair plus a gluon. One can trace on Feynman graphs that this is exactly
the same mechanism of gluon shadowing as glue-glue fusion in a different
reference frame.

Note that Eqs.~(\ref{40})-(\ref{50}) assume that for the coherence
length $l_c\gg R_A$. Even if this condition is satisfied for a $c\bar c$
fluctuation, it can be broken for the $c\bar cG$ component which is 
heavier. Indeed, it was found in \cite{KRT} that the coherence length
for gluon shadowing as about an order of magnitude shorter than the one
for shadowing of sea quarks. Therefore, one should not rely on the long 
coherence length approximation used in (\ref{40})-(\ref{50}), but take 
into account the finiteness of $l^G_c$. This can be done by using the 
light-cone Green function approach developed in \cite{KST}.

The factor $R_G(x,Q^2,b)$ has the form,
\BE
  R_G(x,Q^2,b) = 1-\frac{\Delta\sigma(\gamma^*A)}
  {A\,\sigma(\gamma^*N)}~,
  \label{100}
\EE
where $\sigma(\gamma^*N)$ is the part of the total $\gamma^*N$ cross 
section related to a $c\bar c$ fluctuation in the photon,
\BE
  \sigma(\gamma^*N) = \int d^2r_T
  \int\limits_0^1 d\alpha\,
  \left|\Psi^{\gamma^*}_{c\bar c}(r_T,\alpha,Q^2)\right|^2\,
  \sqq(r_T,x)\ .
  \label{105}
\EE
Here $\Psi^{\gamma^*}_{c\bar c}(r_T,\alpha,Q^2)$ is the light-cone wave
function of the $c\bar c$ pair with transverse separation $\vec r_T$ and
relative sharing of the longitudinal momentum $\alpha$ and $1-\alpha$
(see details in \cite{KST}). The numerator $\Delta\sigma(\gamma^*A)$ in
(\ref{100}) reads \cite{KST},
\BA
  \Delta\sigma(\gamma^*A) &=&
    8\pi\,{\rm Re}\int\!d^2b \int\!dM^2\,
      \left.\frac{d^2\sigma(\gamma^*N\to c\bar cGN)}{dM^2\,dq_T^2}
      \right|_{q_T=0}
  \label{110}\\
  &\times&
  \int\limits_{-\infty}^{\infty}\!\!dz_1 
  \int\limits_{-\infty}^{\infty}\!\!dz_2\,
    \Theta(z_2-z_1)\,\rho_A(b,z_1)\,\rho_A(b,z_2)\,
    \exp\left[-i\,q_L\,(z_2-z_1)\right]~.
  \nonumber
\EA
Here the invariant mass squared of the $c\bar cG$ system reads,
\BE 
  M^2=\sum\limits_i\frac{m_i^2+k_i^2}{\alpha_i}\ ,
  \label{115}
\EE
where the sum is taken over partons ($c\bar cG$) having mass $m_i$,
transverse momentum $\vec k_i$ and fraction $\alpha_i$ of the full
momentum. The $c\bar cG$ system is produced diffractively as an
intermediate state in a double interaction in the nucleus. $z_1$ and
$z_2$ are the longitudinal coordinates of the nucleons $N_1$ and $N_2$,
respectively, participating in the diffractive transition $\gamma^*\,N_1
\to c\bar cG\,N_1$ and back $c\bar cG\,N_2\to\gamma^*\,N_2$. The value
of $\Delta\sigma$ is controlled by the longitudinal momentum transfer
\BE
  q_L=\frac{Q^2+M^2}{2\,\nu}~,
  \label{120}
\EE
which is related to the gluonic coherence length $l^G_c=1/q_L$.
 
The propagation and interaction of the $c\bar cG$ system in the nuclear 
medium between the points $z_1$ and $z_2$ is described by the Green
function $G_{c\bar cG}(\vec r_2,\vec\rho_2,z_2;\vec r_1,\vec\rho_1z_1)$,
where $\vec r_{1,2}$ and $\vec\rho_{1,2}$ are the transverse separations 
between the $c$ and $\bar c$ and between the $c\bar c$ pair and gluon
at the point $z_1$ and destination $z_2$ respectively. Then the Fourier
transform of the diffractive cross section in \Ref{110}
\BE
  8\pi\int\!dM_X^2\,
  \left.\frac{d^2\sigma(\gamma^*N\to XN)}{dM_X^2\,dq_T^2}
  \right|_{q_T=0}\!\!\!\cos\left[q_L\,(z_2-z_1)\right]
\EE
can be represented in the form,
\BA
  &&{1\over2}\,{\rm Re}\int\!d^2\!r_2 d^2\!\rho_2 d^2\!r_1 d^2\!\rho_1
  \int d\alpha_q d\ln(\alpha_G)
  \label{130}\\
  &&\times\,
  F^{\dagger}_{\gamma^*\to c\bar cG}(\vec r_2,\vec\rho_2,\alpha_q,\alpha_G)~
  G_{c\bar cG}(\vec r_2,\vec\rho_2,z_2;\vec r_1,\vec\rho_1,z_1)~
  F_{\gamma^*\to c\bar cG}(\vec r_1,\vec\rho_1,\alpha_q,\alpha_G)~.
  \nonumber
\EA

Assuming that the momentum fraction taken by the gluon is small,
$\alpha\ll 1$, and neglecting the $c\bar c$ separation $r\ll\rho$
we arrive at a factorized form of the three-body Green function,
\BE
  G_{c\bar cG}(\vec r_2,\vec\rho_2,z_2;\vec r_1,\vec\rho_1,z_1)
  \Rightarrow
  G_{c\bar c}(\vec r_2,z_2;\vec r_1,z_1)\;
  G_{GG}(\vec\rho_2,z_2;\vec\rho_1,z_1)~,
  \label{140}
\EE
where $G_{GG}(\vec\rho_2,z_2;\vec\rho_1,z_1)$ describes propagation of the
$GG$ dipole (in fact the color-octet $c\bar c$ and gluon) in the nuclear
medium. This Green function satisfies the two dimensional Schr\"odinger
equation which includes the glue-glue nonperturbative interaction via the
light-cone potential $V(\vec\rho,z)$, as well as interaction with the
nuclear medium.
\BE
  i\,\frac{d}{dz_2} G_{GG}(\vec\rho_2,z_2;\vec\rho_1,z_1) =
  \left[-\frac{\Delta(\vec\rho_2)}{2\,\nu\,\alpha_G(1-\alpha_G)}
  +V(\vec\rho_2,z_2)\right]\,G_{GG}(\vec\rho_2,z_2;\vec\rho_1,z_1)~,
  \label{150}
\EE
where
\BE
  2\,{\rm Im}\,V(\vec\rho,z) =
  -\sigma_{GG}(\vec\rho)\,\rho_A(b,z)~,
  \label{160}
\EE
and the glue-glue dipole cross section is related to the $q\bar q$ one
by the relation,
\BE 
  \sigma_{GG}(r,x)={9\over4}\,\sqq(r,x)~. 
  \label{180}
\EE 

Following \cite{KST} we assume that the real part of the potential
has a form
\BE
  {\rm Re}\,V(\vec \rho,z) =
  \frac{b_0^4\,\rho^2}{2\,\nu\,\alpha_G(1-\alpha_G)}~.
  \label{170}
\EE
The parameter $b_0=0.65\GeV$ was fixed by the data on diffractive gluon 
radiation (the triple-Pomeron contribution in terms of Regge approach)
which is an essential part of Gribov's inelastic shadowing \cite{Gribov}.
The well known smallness of such a diffractive cross section explains why
$b_0$ is so large, leading to a rather weak gluon shadowing. In other words,
this strong interaction squeezes the glue-glue wave packet resulting in
small nuclear attenuation due to color transparency. 
 
In order to get an analytical solution of Eq.~(\ref{160}) we assume that
$\sigma_{GG}(r,s)\approx C_{GG}(s)\,r^2$, with $C_{GG}(s) = \left.
d\sigma_{GG}(r,s)/dr^2\right|_{r=0}$. We also use the approximation
of constant nuclear density $\rho_A(r)=\rho_0\,\Theta(R_A-r)$ which
is rather accurate for heavy nuclei. Correspondingly $T_A(b)=\rho_0\,L(b)$,
where $L(b)=2\sqrt{R_A^2-b^2}$.

With these simplifications we can perform integration over $z_2$ in
\Ref{110} and the gluon shadowing correction takes the form,
\BE
  \label{RGdef}
  1-R_G(x,Q^2) = \rho_0^2\int \frac{d^2b}{T_A(b)}
  \int\limits_0^{L(b)}\!\! dz \,\left(L(b)-z\right) 
  W(x,Q^2,b,z)~,
\EE
where $W(x,Q^2,b,z)$ reads \cite{KST},
\BA
  \label{Wdef}
  W(x,Q^2,b,z) &=&
    \frac{4 \alpha_s(\tQ^2)}{3 \pi C_{qq}(x)}\,
    \int\limits_x^{x_\max}\!\!d\alpha_G
    \frac{C_{GG}^2(\tx,Q^2,b)}{\alpha_G{\tb}^2}\,
    \mbox{Re}\left\{
      \vphantom{\frac{1^1}1}
      \exp\left(-\frac{\Omega z}{2 t}\right)
    \right. \nonumber\\
    &\times&\!\!\!\!\!\left.\left[
    \frac tw + \frac{\sinh \left(\Omega z\right)}t\,
    \ln\!\left(1-\frac{t^2}{u^2}\right) +
    \frac{2 t^3}{u w^2} +
    \frac{t \sinh\!\left(\Omega z\right)}{w^2} +
    \frac{4 t^3}{w^3}
    \right]\right\},\\
    \alpha_s(Q^2) &=& \frac{4\pi}{9\ln\left[
       \left(Q^2+0.25\,\GeV^2\right)/\Lambda^2
    \right]}\,,\\
    \tx    &=& \mbox{min}\left(x/\alpha_G,x_\max\right) \,,\\
    \tQ^2  &=& Q^2 + M_\Psi^2                           \,,\\        
    \tb^2  &=& \tb_0^2 + \alpha_G\,\tQ^2                \,,\\
    B      &=& \sqrt{\tb^4 - i\alpha_G(1\!-\!\alpha_G)%
               \, \nu\, \rho_0\, C_{GG}(\tx,b)}         \,,\\
    \Omega &=& i B/\left(\alpha_G(1\!-\!\alpha_G) \nu \right) \,,\\
    t &=& B\,/\,\tb^2                                                   \,,\\
    u &=&    t   \cosh\left(\Omega z\right)+  \sinh\left(\Omega z\right)\,,\\
    w &=& (1+t^2)\sinh\left(\Omega z\right)+2t\cosh\left(\Omega z\right)\,.
 \EA
Here $x_\max = 0.1$ and $\Lambda = 200\MeV$. The factor $C_{GG}(x,Q^2,b)$
is determined by the condition of equivalent descriptions in the limit
$l^G_c\gg R_A$ using the realistic parameterization for $\sigma_{GG}(r_T)$
and its simplified form $\sigma_{GG}(r_T)=C_{GG}r_T^2$,
\BE
  \label{CGG}
  \left<\!\!\left<\vphantom{\frac94} C_{GG}(x,Q^2,b)
  \,r_T^2 \,T_A(b)\right>\!\!\right> = 
  \left<\!\!\left<\frac94\sigma_{q\bar q}(x,r_T)
  \,R_G(x,Q^2,b)\,T_A(b)\right>\!\!\right>\,.
\EE
Here
\BE
  \left<\hskip-3pt\left< f(r_T) \right>\hskip-3pt\right>\,\, \equiv\,
  \frac{\int\!d^2\vrT |\Psi_{qG}(r_T)|^2_\eff 
  \left(1-e^{-\frac12 f(r_T)}\right)}
  {\int\!d^2\vrT |\Psi_{qG}(r_T)|^2_\eff \, f(r_T)}\,,
\EE
and the nonperturbative light-cone wave function of the $GG$ (in fact
$c\bar c$ and $G$) dipole, 
\BE
  |\Psi_{GG}(r_T)|^2_\eff \propto \frac{e^{-b_0^2\,r_T^2}}{r_T^2}\,.
\EE

Obtaining the gluon shadowing factor $R_G(b)$ one can relate it to the
corresponding value $T_A(b)$. This gives $R_G$ as a function of thickness
$T_A$, $R_G(T_A)$. This function is then used in calculations where
$T_A(b)$ is obtained with the realistic Woods-Saxon nucleus density.

Fig.~\ref{SG-all} shows the ratios of cross sections calculated with
and without gluon shadowing
\BE
  S_G(s,Q^2) = \frac{\sigma^{\gamma^*A}_G(s,Q^2)}
                    {\sigma^{\gamma^*A}(s,Q^2)}
\EE
for incoherent and coherent exclusive charmonium electroproduction.
\BF
\centerline{
  \PSfig{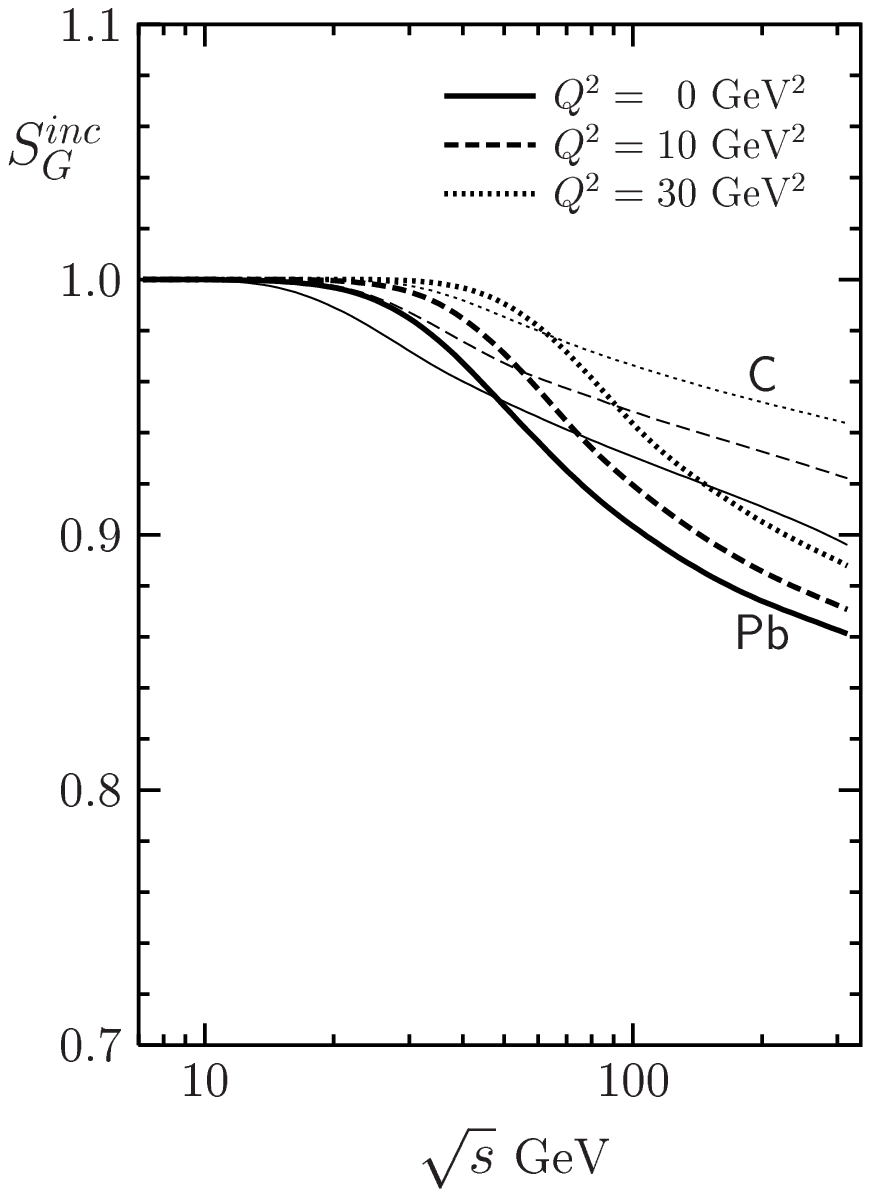}\hskip5mm
  \PSfig{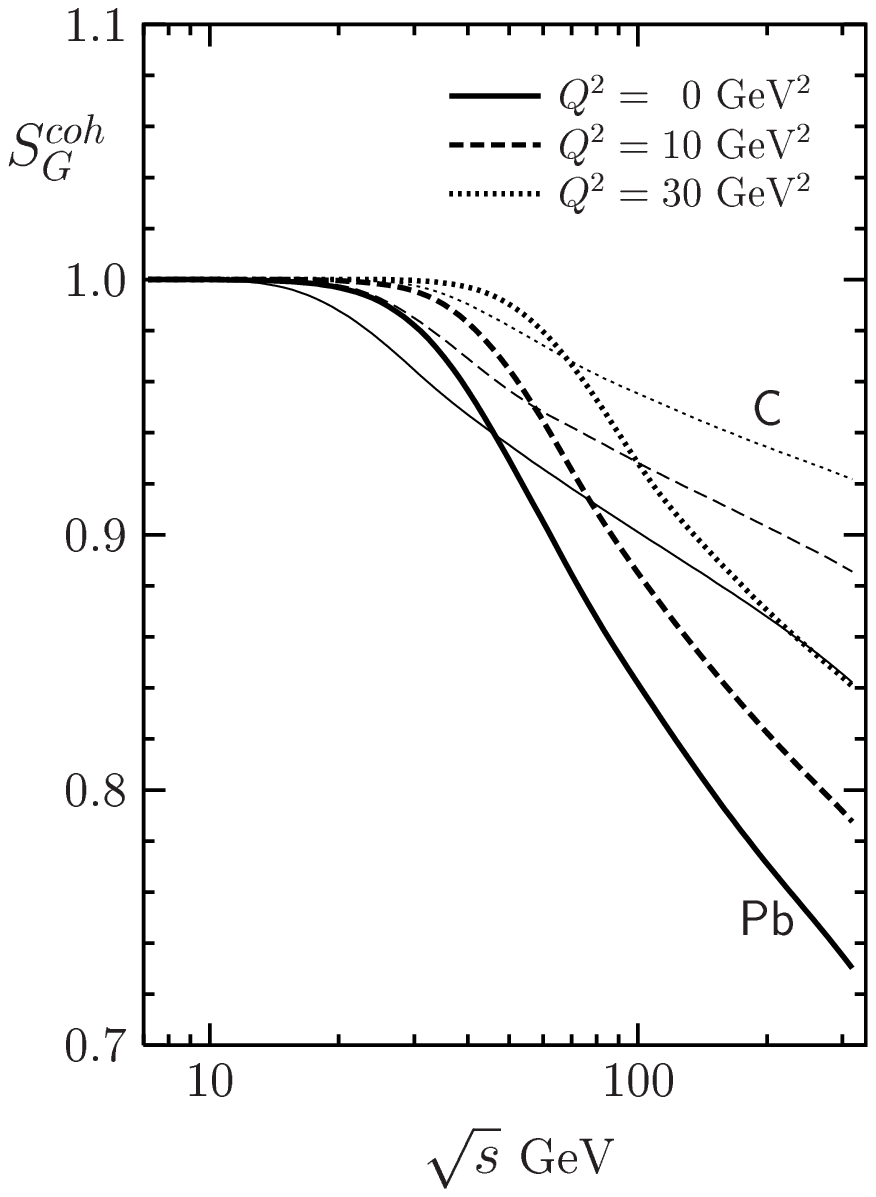}
}
\Caption{
  Ratios $S_G(s,Q^2)$ of cross sections calculated with and without
  gluon shadowing for incoherent and coherent charmonia production.
  We only plot ratios for $\Jpsi$ production, because ratios for
  $\psi'$ production are practically the same. All curves are
  calculated with the GBW parameterization of the dipole cross
  section $\sqq$.
  \label{SG-all}
}
\EF
We see that the onset of gluon shadowing happens at a c.m. energy of few
tens GeV. This is controlled by the nuclear formfactor which depends on
the longitudinal momentum transfer $q_c=1/l^G_c$. It was found in 
\cite{KRT} that the coherence length for gluon shadowing is rather 
short,
\BE
  l^G_c \approx \frac{1}{10\,x\,m_N}~,
  \label{190}
\EE
where $x$ in our case should be an effective one, $x=(Q^2+M_\Psi^2)/2\nu$.
The onset of shadowing should be expected at $q_c^2\sim 3/(R_A^{ch})^2$
corresponding to
\BE
  s_G \sim 10 m_N R_A^{ch}(Q^2+M_\Psi^2)/\sqrt{3}~,
\EE
where $(R_A^{ch})^2$ is the mean square of the nuclear charge radius.
This estimate is in a good agreement with Fig.~\ref{SG-all}. Remarkably,
the onset of shadowing is delayed with rising nuclear radius and $Q^2$.

Fig.~\ref{kT-gl} shows the effect of shadowing for the $k_T$ dependence
of the coherent photoproduction. As one can see these corrections are
quite small for $Q^2=0$ (and even smaller with increasing $Q^2$, see
Eqs. \Ref{RGdef} and \Ref{Wdef}).
\BF
\centerline{
  \PSfig{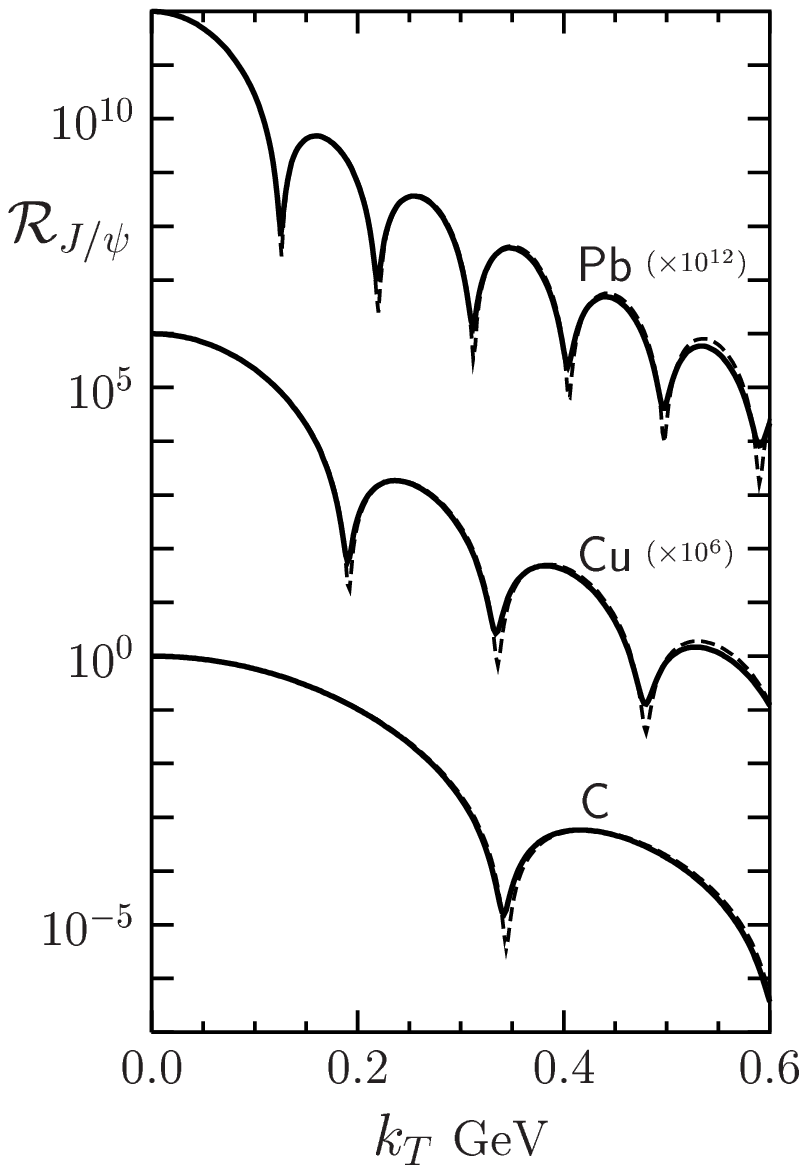}\hskip5mm
  \PSfig{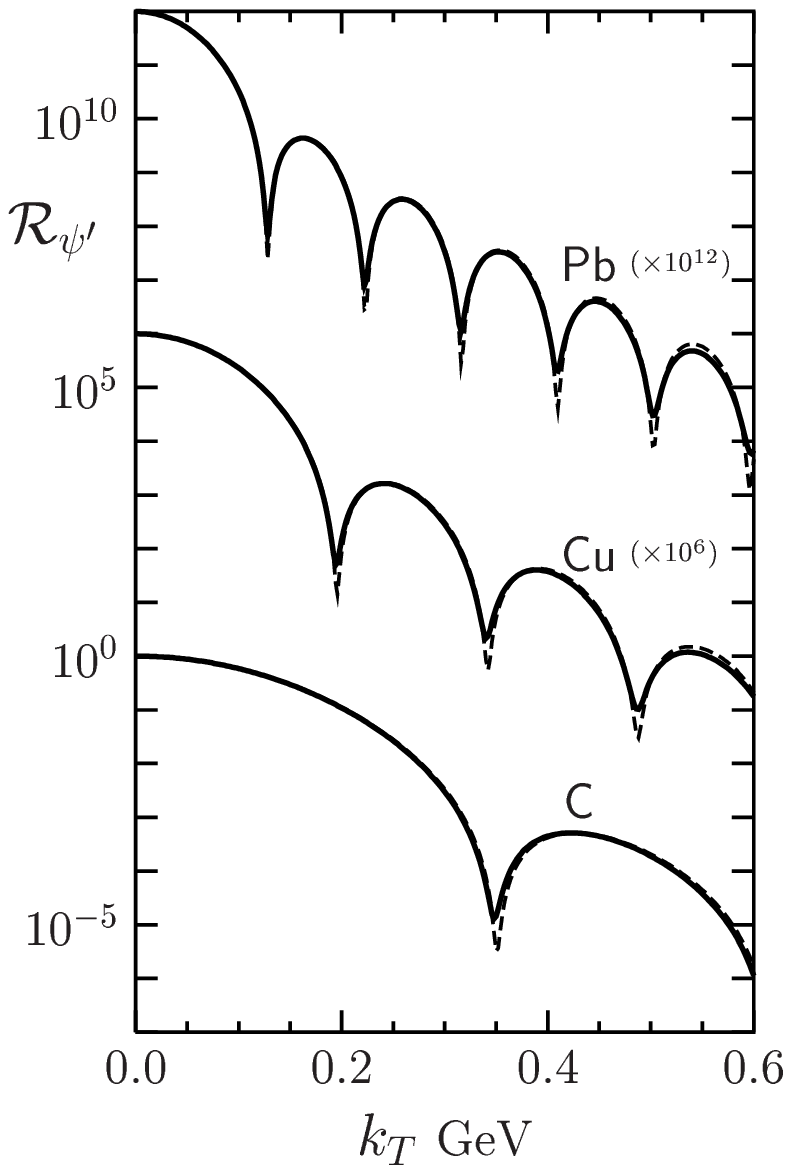}
}
\Caption{
  \label{kT-gl}
  Ratio ${\cal R}_\Psi$ as a function of $k_T$ at $s=4000\GeV^2$
  and $Q^2=0$ with gluon shadowing (solid curves) and without (dashed
  curves) with the GBW parameterization of the dipole cross section.
}
\EF

\section{Finite coherence length. \label{section-lc}}

A strictly quantum-mechanical treatment of a fluctuating $q\bar q$ pair
propagating through an absorptive medium based on the LC Green function
approach has been suggested recently in \cite{KNST}. However, an
analytical solution for the LC Green function is known only for the
simplest form of the dipole cross section $\sqq(r_T) \propto r_T^2$. With
a realistic form of $\sqq(r_T)$ it is possible only to solve this problem
numerically, what is still a challenge. Here we use the approximation
suggested in \cite{HKZ} to evaluate the corrections arising from the
finiteness of $l_c=2\nu/M_{\Psi}^2$ by multiplying the cross sections
for incoherent and coherent production evaluated for $l_c\to\infty$ by
a kind of formfactor $F^{inc}$ and $F^{coh}$ respectively:
\BA
  \sigma^{\gamma A\to\Psi X}(s,Q^2) &\Rightarrow&
  \sigma^{\gamma A\to\Psi X}(s,Q^2) \cdot F^{inc}\!
  \left(s,l_c(s,Q^2)\right)\,,\\
  \sigma^{\gamma A\to\Psi A}(s,Q^2) &\Rightarrow&
  \sigma^{\gamma A\to\Psi A}(s,Q^2) \cdot F^{coh}\!
\left(s,l_c(s,Q^2)\right)\,,
\EA
where
\BA
  F^{inc}(s,l_c)   &=& \int\!d^2\bb
    \,\int\limits_{-\infty}^\infty\!dz\,\rho_A(\bb,z)
    \,\left|F_1(s,\bb,z)-F_2(s,\bb,z,l_c)\right|^2 / \,
    (...)|_{l_c = \infty} \,,
  \label{200}\\
  F^{coh}(s,l_c)   &=& \int\!d^2\bb
    \left|
    \,\int\limits_{-\infty}^\infty\!dz\,\rho_A(\bb,z)
    \,F_1(s,\bb,z)\,e^{iz/l_c}
    \right|^2 / \,(...)|_{l_c = \infty} \,,
  \label{210}\\
  F_1(s,\bb,z)     &=&
    \,\exp\left(-\frac12\,\sigma_{\Psi N}(s)
    \!\int\limits_z^\infty\!dz'\,\rho_A(\bb,z')\right)\,,
  \label{220}\\
  F_2(s,\bb,z,l_c) &=& \frac12\,\sigma_{\Psi N}(s)
    \!\int\limits_{-\infty}^z\!dz'
    \,\rho_A(\bb,z')\,F_1(\bb,z')\,e^{-i(z-z')/l_c}\,.
  \label{230}
\EA
For the charmonium nucleon total cross section $\sigma_{\Psi N}(s)$ we use
our previous results \cite{HIKT}, essentially the expectation value of
the dipole cross section $\sqq(r_T,s)$ taken between the LC wave functions
for the charmonia. The ``formfactors'' $F^{inc}$ for incoherent 
$\Jpsi$ and $\psi'$ production are depicted in Fig.~\ref{FigFI}.
\BF
\centerline{
  \PSfig{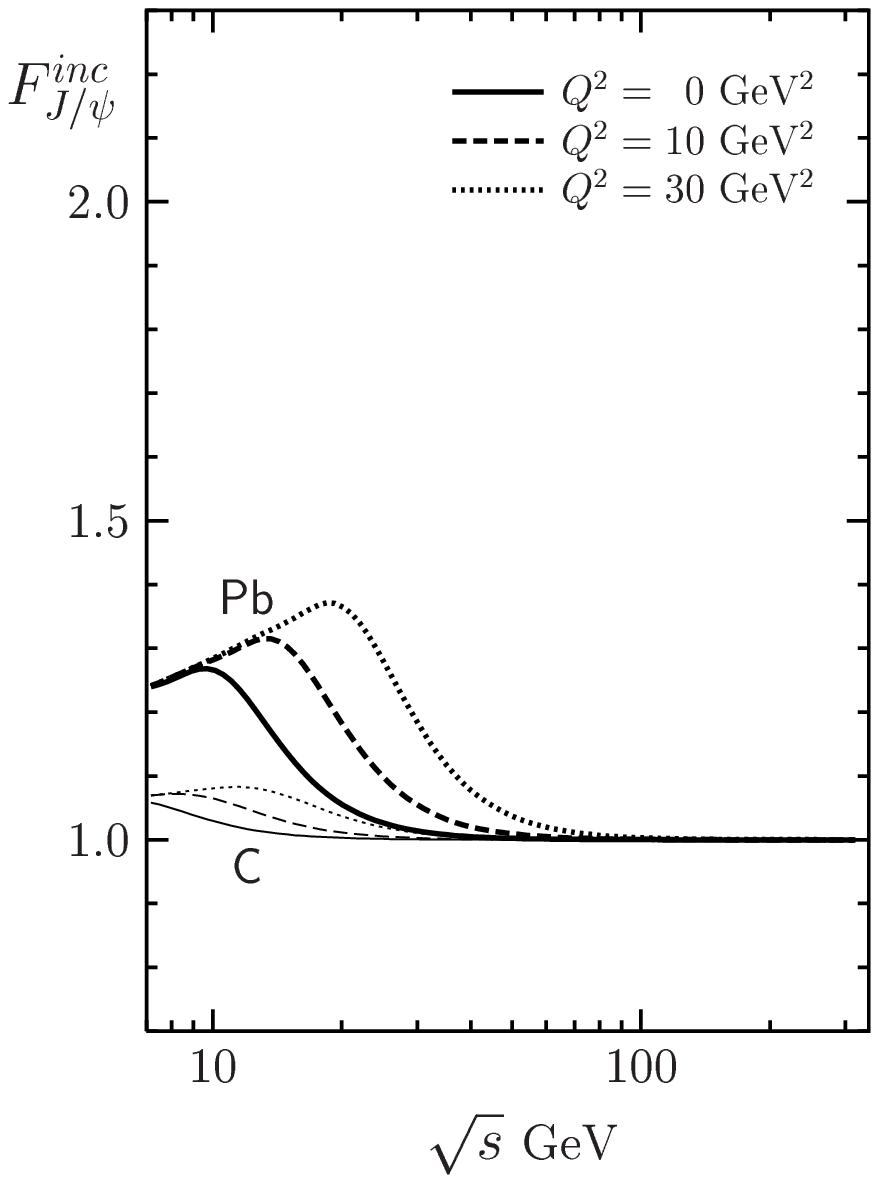}\hskip5mm
  \PSfig{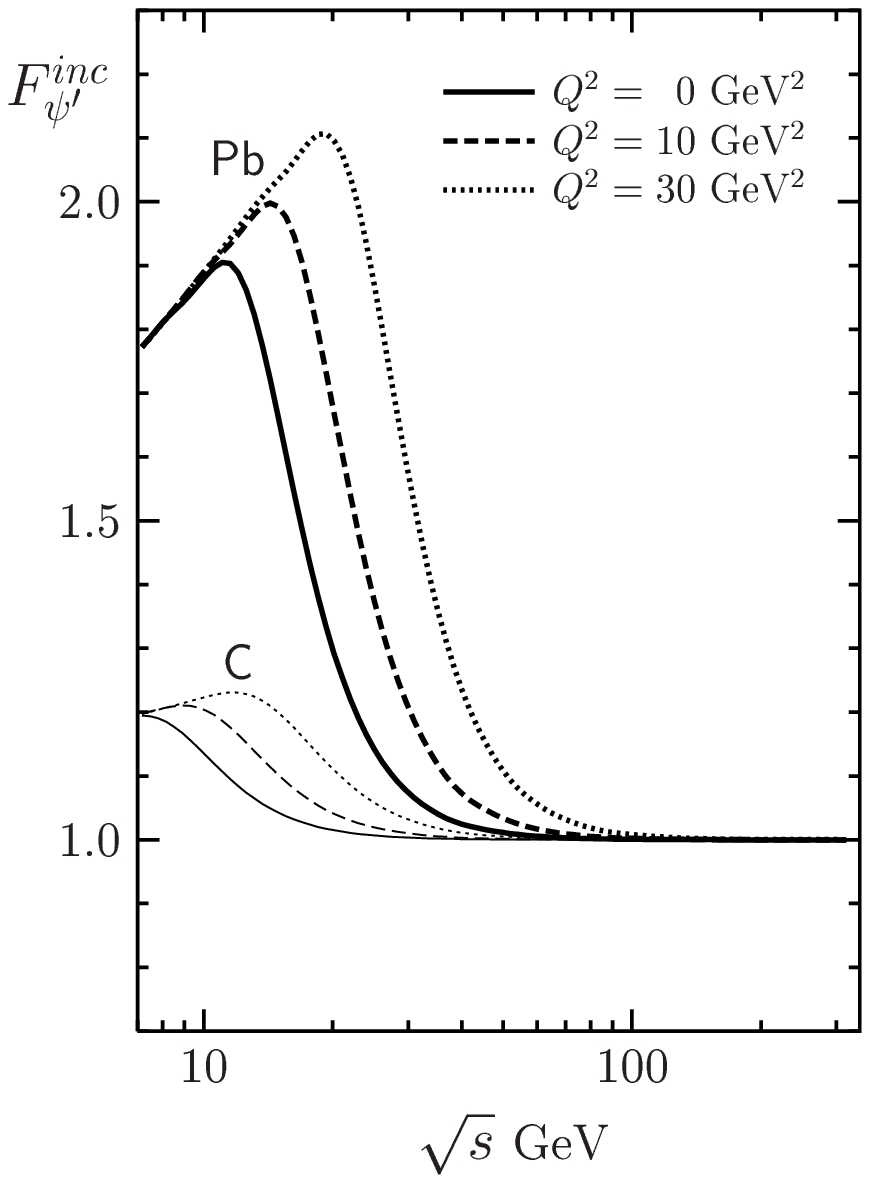}
}
\Caption{
  \label{FigFI}
  $l_c$ corrections for incoherent production of $\Jpsi$ and
  $\psi'$ on C (thin curves) and Pb (thick curves) for the GBW 
  parameterization of the dipole cross section $\sqq$. }
\EF
The observed nontrivial energy dependence is easy to interpret. At low
energies $l_c \ll R_A$ and the photon propagates without any attenuation
inside the nucleus where it develops for a short time $t_c$ a $c\bar c$
fluctuation which momentarily interacts to get on mass shell. The produced
$c\bar c$ pair attenuates along the path whose length is a half of the nuclear
thickness on the average. On the other hand, at high energies $l_c\gg R_A$
and the $c\bar c$ fluctuation is developed long before its interaction with
the nucleus. As a result, it propagates through the whole nucleus and the
mean path length is twice as long as at low energies. This is why the nuclear 
transparency drops when going from the regimes of short to long $l_c$. 

One can simplify expression (\ref{200}) assuming $\sigma_{\Psi N}T_A(b)\ll 1$
which is a rather accurate approximation for $\Jpsi$. Then one can expand the
exponentials in (\ref{220})-(\ref{230}) and obtain \cite{BKMNZ},
\BE
  F^{inc}_{\Jpsi}(s) = 
  \frac{1-{1\over2}\Bigl[1+F^2_A(q_c)\Bigr]\sigma_{\Jpsi N}(s)\la T_A\ra}
  {1-\sigma_{\Jpsi N}(s)\la T_A\ra}~.
  \label{240}
\EE
Here
\BE
  F_A^2(q) = \frac{1}{\la T_A\ra}\int d^2b
  \left|\int\limits_{-\infty}^{\infty} dz\,
  \rho_A(b,z)\,e^{iqz}\right|^2
  \label{260}
\EE
is the so called longitudinal nuclear formfactor.

At low energies, $q_c\equiv1/l_c\gg 1/R_A$, the formfactor $F_A^2(q_c)=0$
and both the numerator and denominator in (\ref{240}) decrease with
energy due to the rising $\sigma_{\Jpsi N}(s)$, but the denominator does
it faster. This is why the ratio Eq.~(\ref{240}) rises with energy unless
it drops due to the rise of the formfactor $F_A^2(q_c) \neq 0$. The
transition energy regime is related to the variation of the formfactor
between 0 and 1, i.e. at 
\BE
  s_{c\bar c} \sim m_N (M_{\Jpsi}^2+Q^2) R^{ch}_A/\sqrt{3}
  = {1\over10} s_G~.
\EE
This energy increases with $Q^2$ and nuclear radius in accordance with 
Fig.~\ref{FigFI}.

\BF
\centerline{
  \PSfig{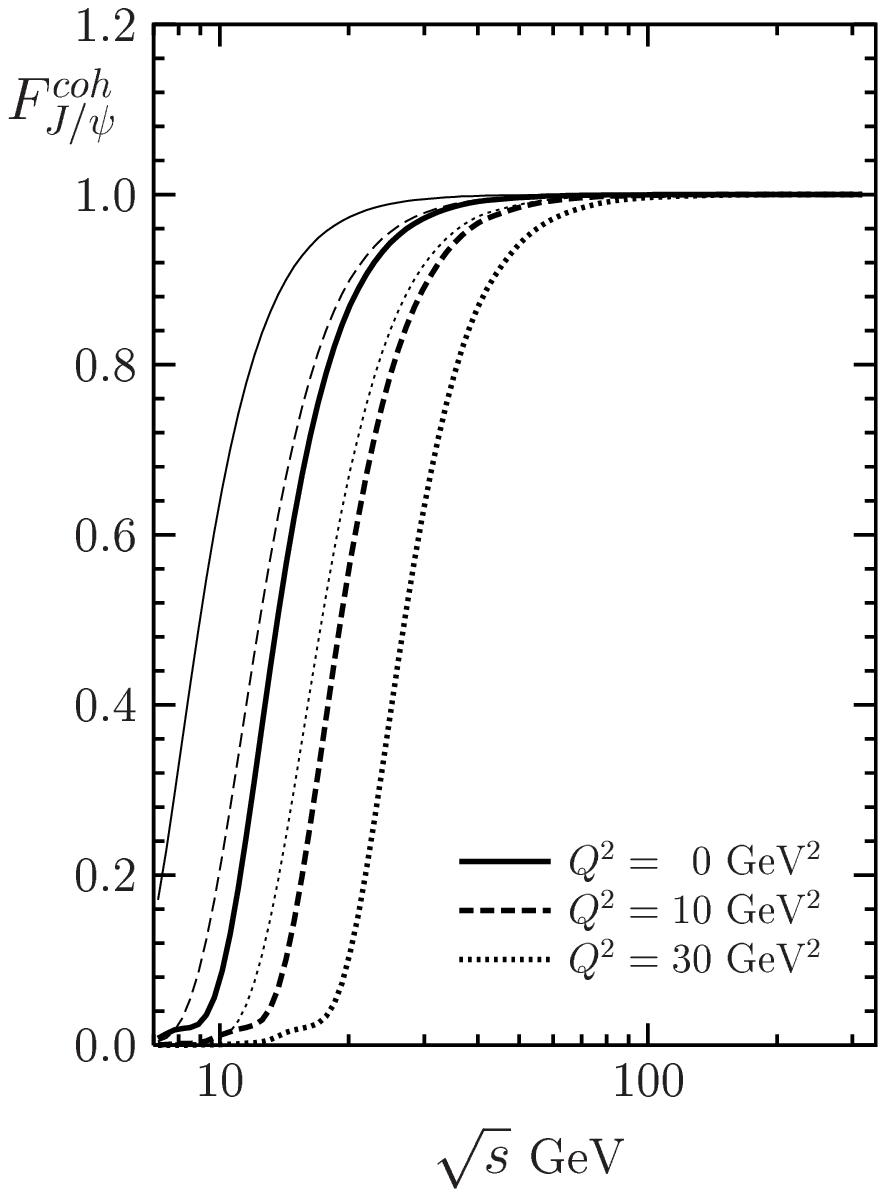}\hskip5mm
  \PSfig{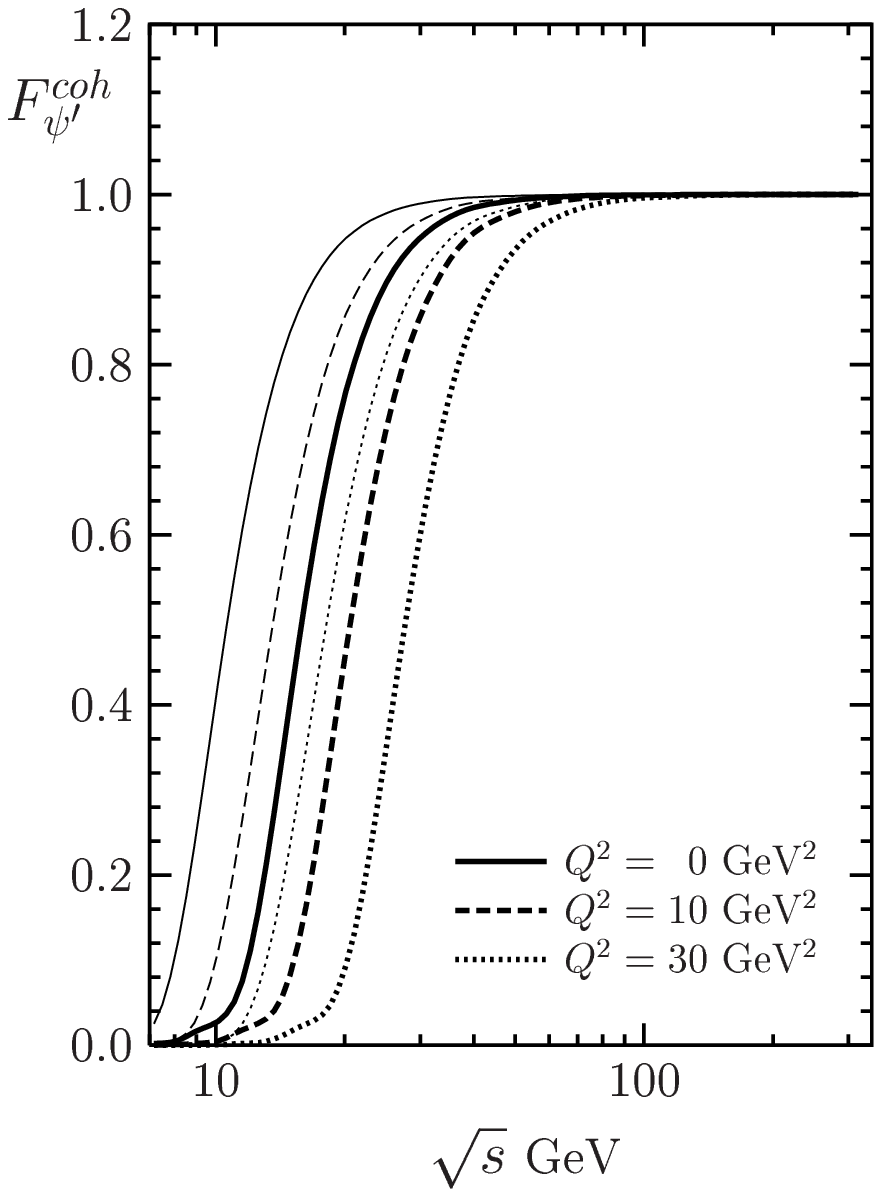}
}
\Caption{
  \label{FigFC}
  $l_c$ corrections for coherent production of $\Jpsi$ and
  $\psi'$ on C (thin curves) and Pb (thick curves) for the GBW
  parameterization of the dipole cross section. Contrary to the
  case of incoherent production results for $\Jpsi$ and $\psi'$
  are very similar.
}
\EF

Note that the Glauber approximation used to calculate $F^{inc}$ is 
rather good for $\Jpsi$, but not for $\psi'$. One could improve it 
applying the rigorous description based on the light-cone Green function 
formalism developed in \cite{KNST} for exclusive production of vector 
mesons. However, it was done only for the unrealistic form of the dipole
cross sections $\sqq \propto r_T^2$ and for an oscillatory wave function
of the vector meson. It is still a challenge to make this approach realistic.

The results for the energy dependence of the formfactor $F^{coh}$ for
coherent scattering are shown on Fig.~\ref{FigFC}. In the limit of low
energies $l_c \to 0$ the strongly oscillating exponential phase factor
in (\ref{210}) makes the integral very small and thus $F^{coh}_{\Jpsi}
\approx 0$. Then the cross section rises with $l_c$ unless it saturates
at $l_c\gg R_A$ when the phase factor becomes constant. Apparently, this
transition region is shifted to higher energies for larger $Q^2$ and
nuclear radius as is confirmed by the curves in Fig.~\ref{FigFC}.

\section{Final results and conclusions\label{section-final}}

Combining the results of the previous sections (i.e. including finite
coherence length and gluon shadowing) we obtain the final results for
ratios the $R_{\Jpsi}$ and $R_{\psi'}$. Figs.~\ref{s-inc-full} and
\ref{Q-inc-full} show the $s$ and $Q^2$ dependences for incoherent
and Figs.~\ref{s-coh-full} and ~\ref{Q-coh-full} for coherent
production of charmonia.
\BF
\centerline{
  \PSfig{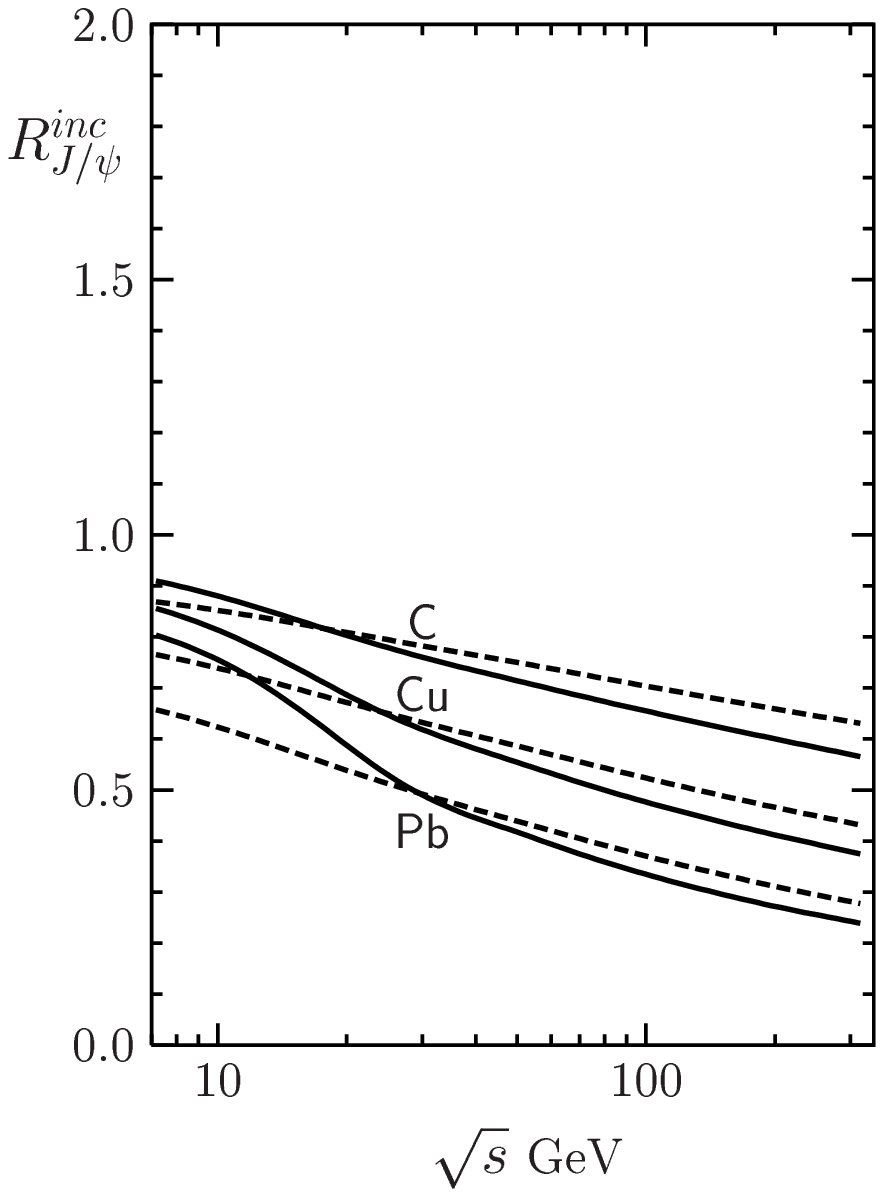}\hskip5mm
  \PSfig{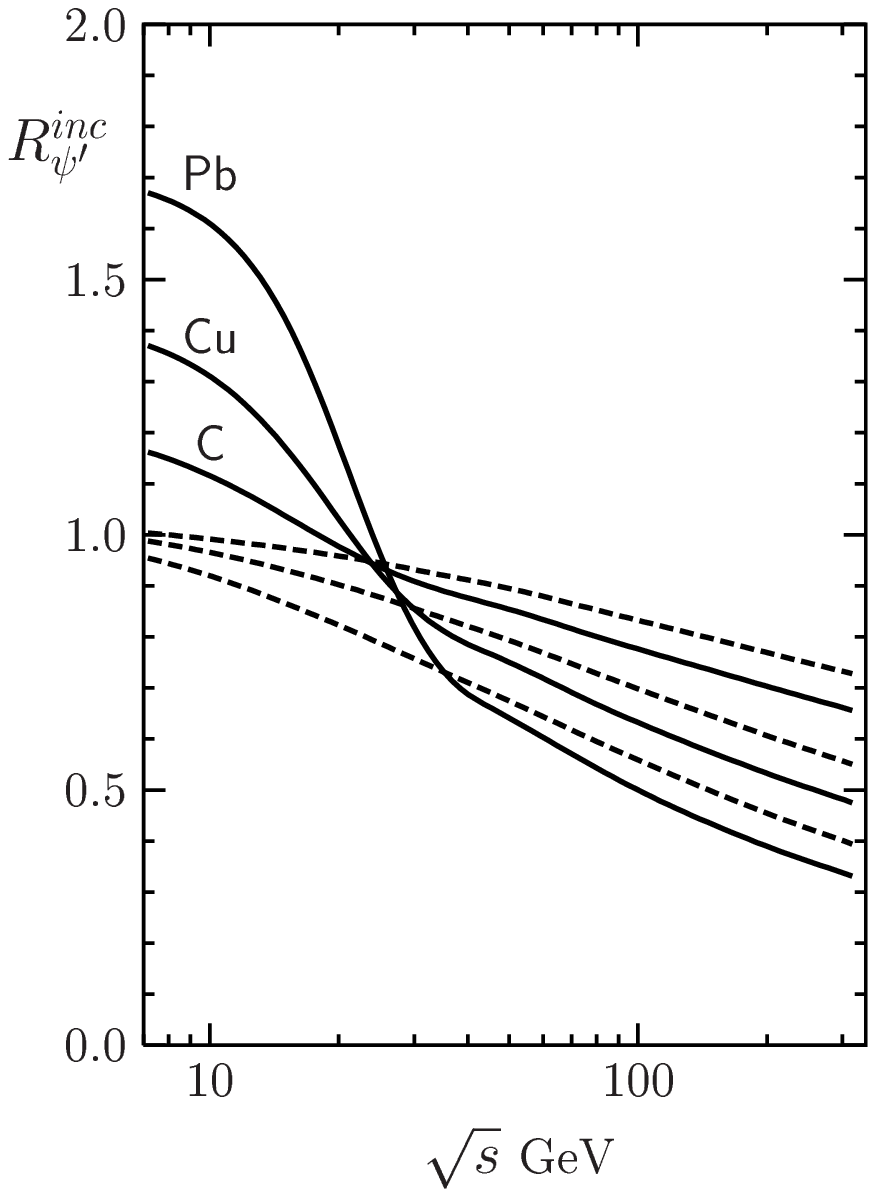}
}
\Caption{
  \label{s-inc-full}
  Ratios $R^{inc}_\Psi$ for $\Jpsi$ and $\psi'$ incoherent production
  on carbon, copper and lead as function of $\sqrt s$ and at $Q^2=0$
  calculated with GBW parameterization of $\sqq$. Solid curves display
  the modifications due to the gluon shadowing and finite coherence
  length $l_c$ while the dashed lines are without (same as on
  Fig.~\ref{s-inc}).
}
\EF
\BF
\centerline{
  \PSfig{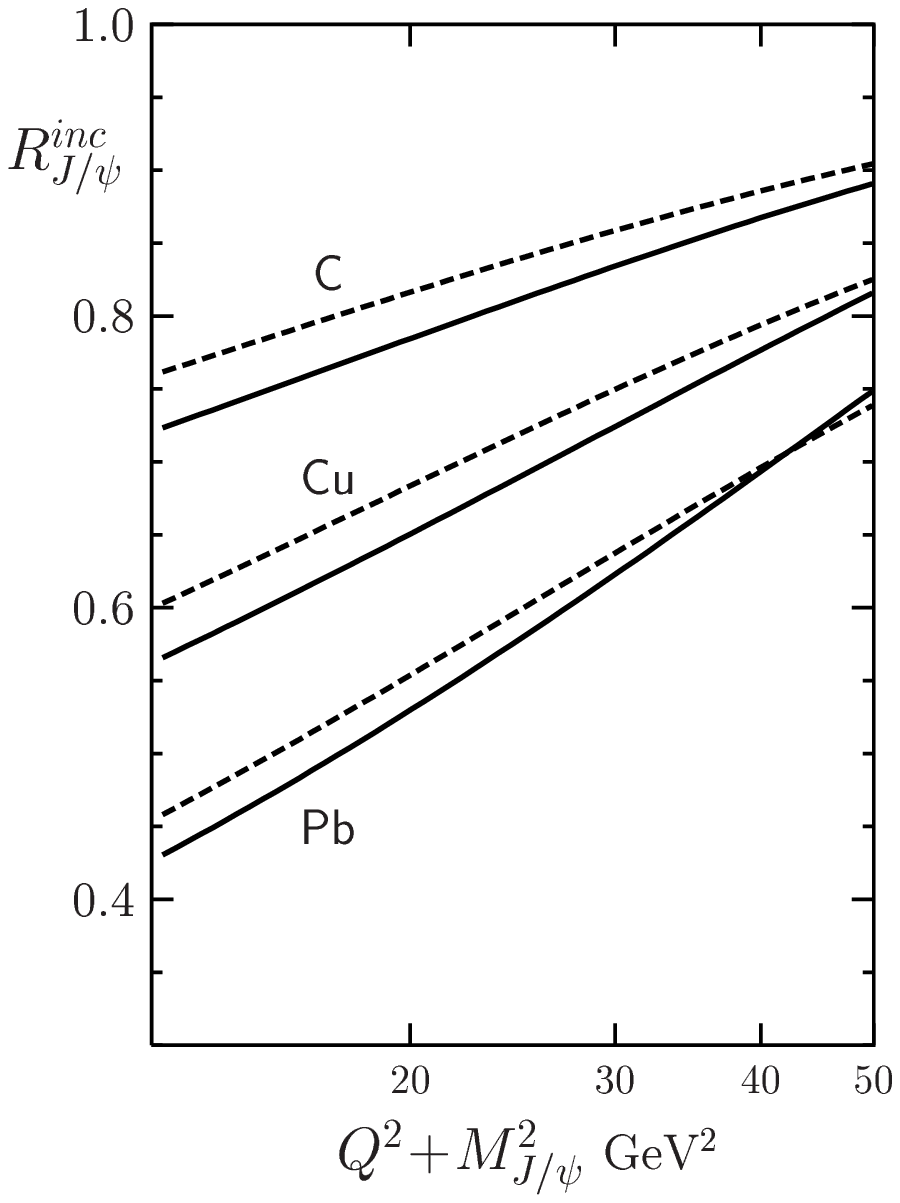}\hskip5mm
  \PSfig{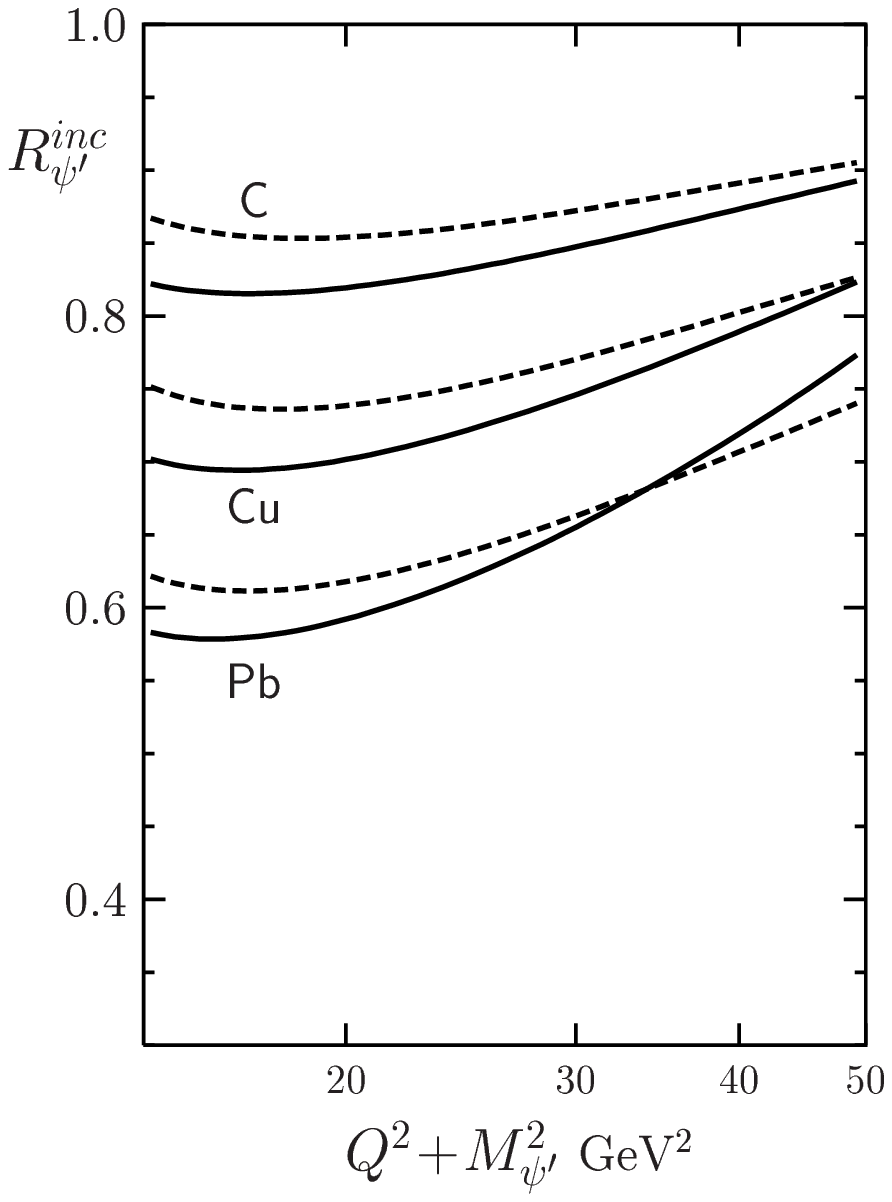}
}
\Caption{
  \label{Q-inc-full}
  Ratios $R^{inc}_\Psi$ for $\Jpsi$ and $\psi'$ incoherent production
  on carbon, copper and lead as function of $Q^2+M^2_\Psi$ at fixed
  $s=4000\GeV^2$ for GBW parameterization of the dipole cross section
  $\sqq$. Solid curves correspond to the final result (with gluon
  shadowing and $l_c$ corrections) while thin curves are without
  (same as on Fig.~\ref{Q-inc}).
}
\EF
\BF
\centerline{
  \PSfig{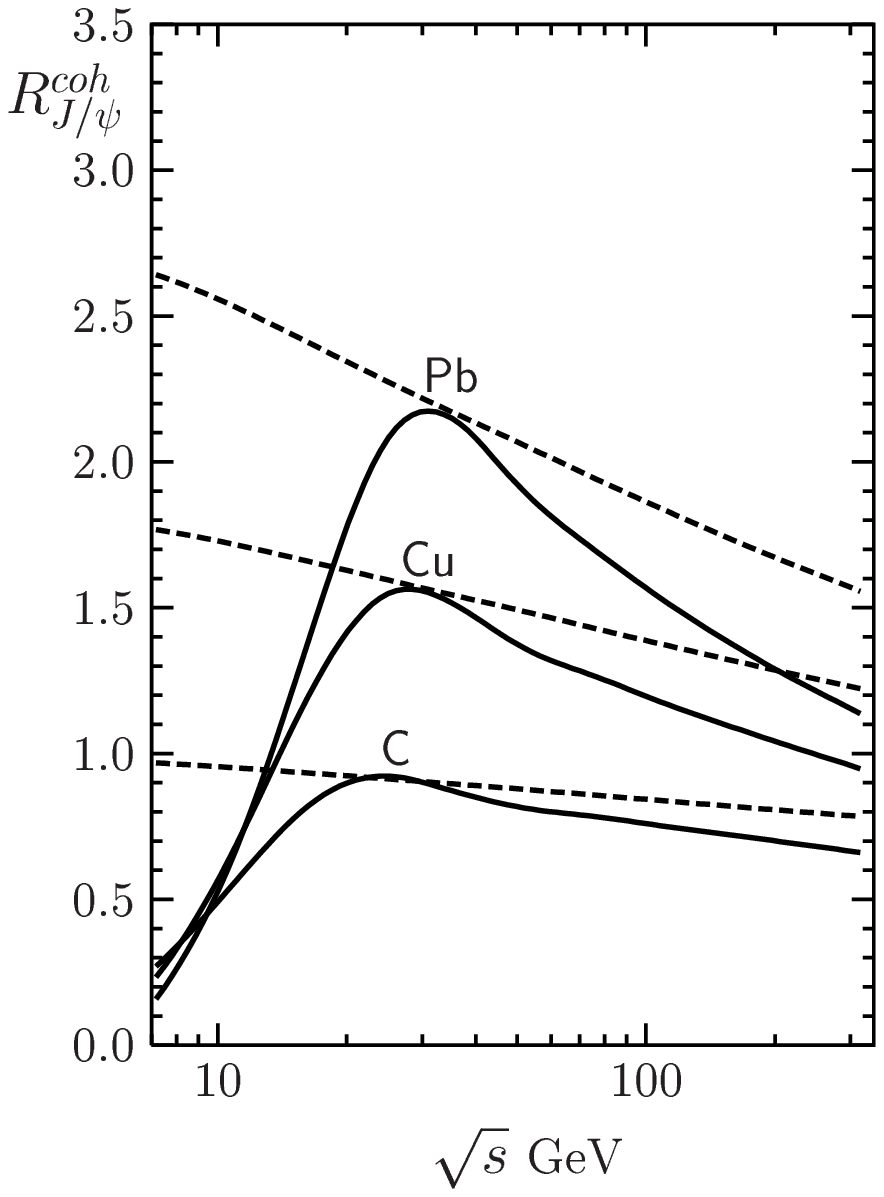}\hskip5mm
  \PSfig{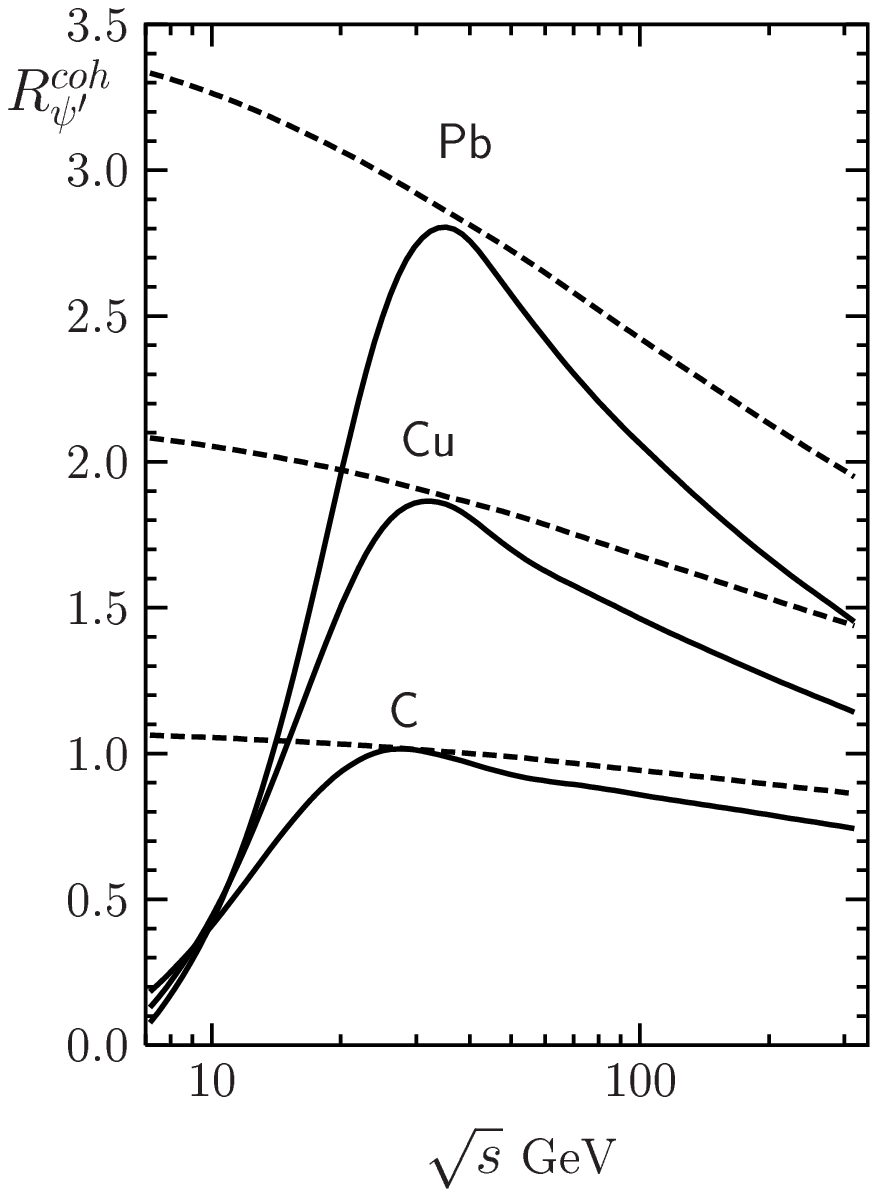}
}
\Caption{
  \label{s-coh-full}
  Ratios $R^{coh}_{\Jpsi}$ and $R^{coh}_{\psi'}$ for coherent production
  on nuclei as a function of $\sqrt s$. The meaning of the different lines
  is the same as in Fig.~\ref{s-inc-full}. Thin curves correspond to 
  Fig.~\ref{s-coh}.
}
\EF
\BF
\centerline{
  \PSfig{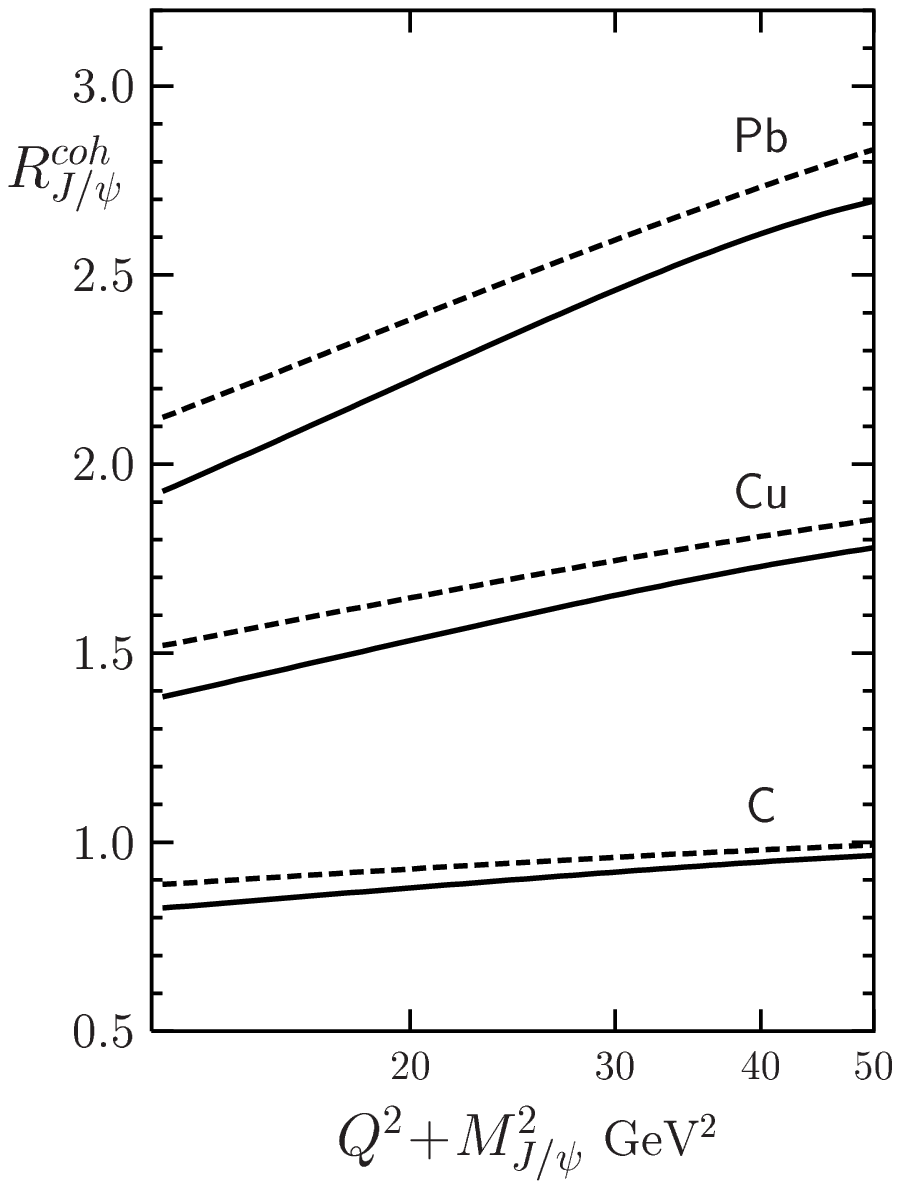}\hskip5mm
  \PSfig{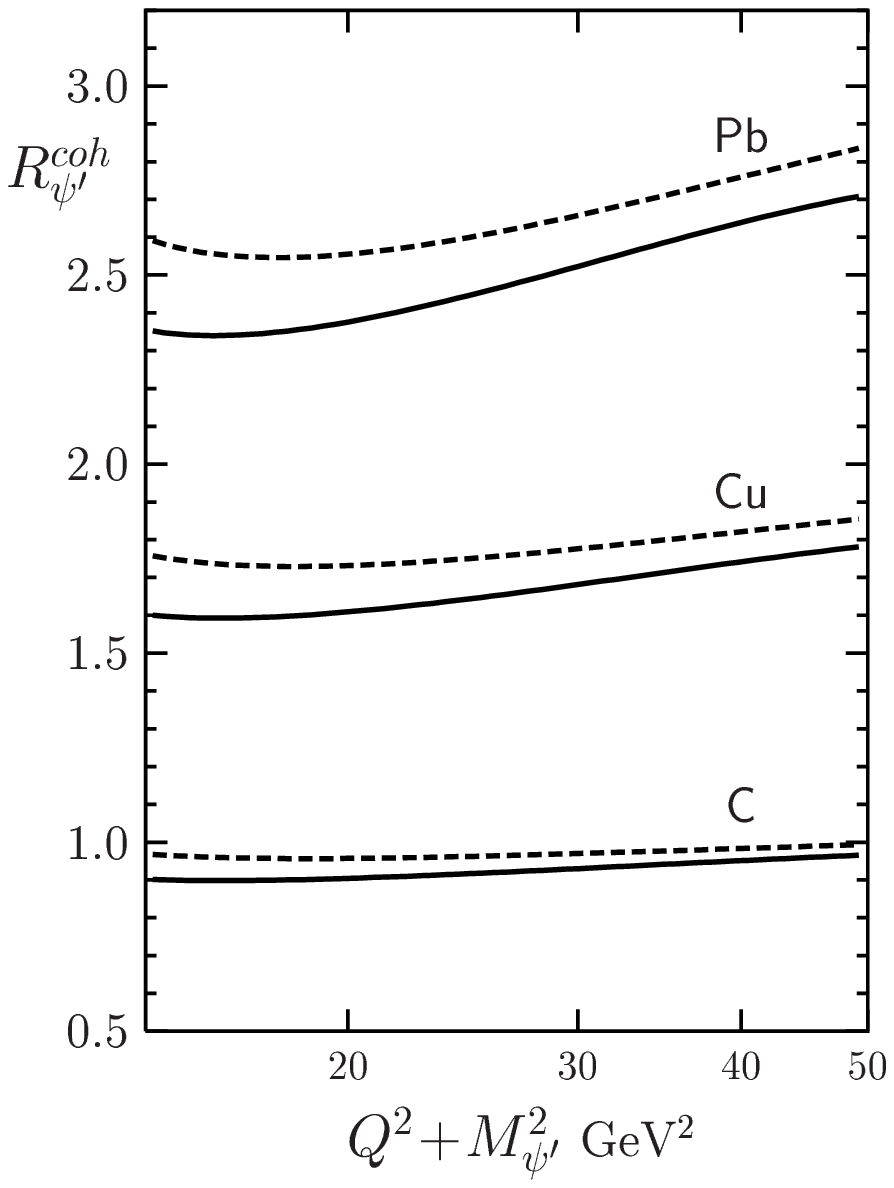}
}
\Caption{
  \label{Q-coh-full}
  Ratios $R^{inc}_\Psi$ for $\Jpsi$ and $\psi'$ coherent production
  on carbon, copper and lead as function of $Q^2+M^2_\Psi$ at fixed
  $s=4000\GeV^2$ for GBW parameterization of the dipole cross section
  $\sqq$. The meaning of the different lines is the same as in
  Fig.~\ref{Q-inc-full}. Thin curves correspond to Fig.~\ref{Q-coh}.
}
\EF

We summarize, in this paper we provide predictions for nuclear effects in
exclusive electroproduction of charmonia on nuclei with realistic
light-cone wave functions of charmonia and with dipole cross sections
which have been already tested in a calculation of elastic charmonium
electroproduction off protons. Such calculations still can be performed
only at rather high energies where the coherence length exceeds the
nuclear size ($l_c\gg R_A$). At these energies gluon shadowing which is
absent at lower energies ($l_c \sim R_A$) becomes an important and
sometimes the dominant effect. Since no reliable experimental
information about gluon shadowing is available so far, we have performed
calculations of this effect employing the same light-cone dipole
approach. We found sizable, but not large corrections (10-20\%)
which, however, keep rising with energy. It turns out that the magnitude
of gluon shadowing is suppressed by a nonperturbative interaction of the
light-cone gluons which is well fixed by data on large mass diffraction.  
The effects related to finiteness of the coherence length at somewhat
lower energies were estimated too.

We have calculated both coherent and incoherent cross sections as function
of energy, $Q^2$ and momentum transfer. Our predictions can be tested in
future experiments at high energies with electron-nuclear colliders (eRHIC).
It also can be produced in peripheral heavy ion collisions (RHIC, LHC),
and our predictions will be published elsewhere.

\noindent {\bf Acknowledgment}: The authors gratefully acknowledge the
partial support by a grant from the Gesellschaft f\"ur Schwerionenforschung
Darmstadt (grant no.~GSI-OR-SCH) and by the Federal Ministry BMBF (grant
no.~06~HD~954). We are grateful to Jan Nemchik who read the manuscript
and made many valuable comments.

\Appendix
\section{Resummation expressions\label{section-resumm}}

Matrix elements of the dipole cross section $\sqq$ for the charmonia
production including the effects of spin rotation were calculated
in \cite{HIKT} for the specific form of the $r_T$ dependence provided
by the KST and GBW parameterizations:
\BE
  \label{dipole-rT}
  \sigma_{q\bar q}(r_T) =
  \sigma_0\left[1-\exp\left(-\frac{r_T^2}{r_0^2}\right)\right]\,.
\EE

In the nuclear case one needs to calculate matrix elements of the operators
with $\sigma_{q\bar q}$ in the exponent. For incoherent and coherent scattering
one can use Taylor expansion, for example:
\BA
  \label{ResumT}
  \left<\!\sigma e^{-c \sigma}\!\right>\!\!
   &=&\!\left<\sigma - c\sigma^2 + \frac{c^2\sigma^3}2   +\cdots\right> \hskip4mm
    = \quad<\!\sigma\!>\cdot\left(1-c f_1+\frac{c^2 f_2}2+\cdots\right),\\
  \label{ResumT-c}
  \left<\!1-e^{-c \sigma}\!\right>\!\!
   &=&\!\left<c\sigma-\frac{c^2\sigma^2}2+\frac{c^3\sigma^3}6 +\cdots\right>
    = c <\!\sigma\!>\cdot\left(1-\frac{c f_1}2+\frac{c^2 f_2}6+\cdots\right),
\EA
where $f_n \equiv <\!\sigma^{n+1}\!>\!/\!<\!\sigma\!>$ and $c$ doesn't depend
on initial and final states. So matrix elements of powers $\sqq^n$ have to be
calculated. The form \Ref{dipole-rT} allows to express any power of $\sqq(r_T)$
as a sum of exponentials $\exp\left(-r_T^2/r_0^2\right)$ with reduced
$r_0^2 \Rightarrow r_0^2/m$, i.e. $<\!\sigma^n\!>$ can be expressed in
terms of $<\!\sigma\!>$ with reduced $r_0^2$.

The rate of convergence of such series slows down with increasing $\sigma$
(larger $s$) and increasing nuclear thickness ($c\propto T_A(b)$). In our
calculations we used a more efficient expansion, which contains matrix
elements $<\!\sigma^n\!>$ in the exponent. This form can be obtained when
applying the identity
\BE
  1+x = e^{\log(1+x)} = \exp\left(x-\frac{x^2}2+\frac{x^3}3+\cdots\right)
\EE
to \Ref{ResumT} (with $x = c f_1+\cdots$) and \Ref{ResumT-c} (with
$x = -c f_1/2+\cdots$). As a result one can get
\BA
  \label{ResumE}
  \left<\sigma e^{-c \sigma}\right> &=&
    <\!\sigma\!> \cdot \exp\left(
      - c f_1 + \frac{c^2}2\left(f_2-f_1^2\right) + \cdots
    \right) \,,\\
  \label{ResumE-c}
  \left<1-e^{-c \sigma}\right> &=& 
    c <\!\sigma\!> \cdot \exp\left(
      - \frac{c f_1}2 + c^2\left(\frac{f_2}6 -\frac{f_1^2}8\right)+\cdots
    \right) \,,
\EA
which converges noticeable faster. Fig.~\ref{FigResum} shows that while
one needs about six terms ($n=6$) to obtain an accurate result using Taylor
expansion \Ref{ResumT}, one has already satisfactory results for $n=2$ in
the method, where one resums in the exponent.
\BF
\centerline{
  \PSfig{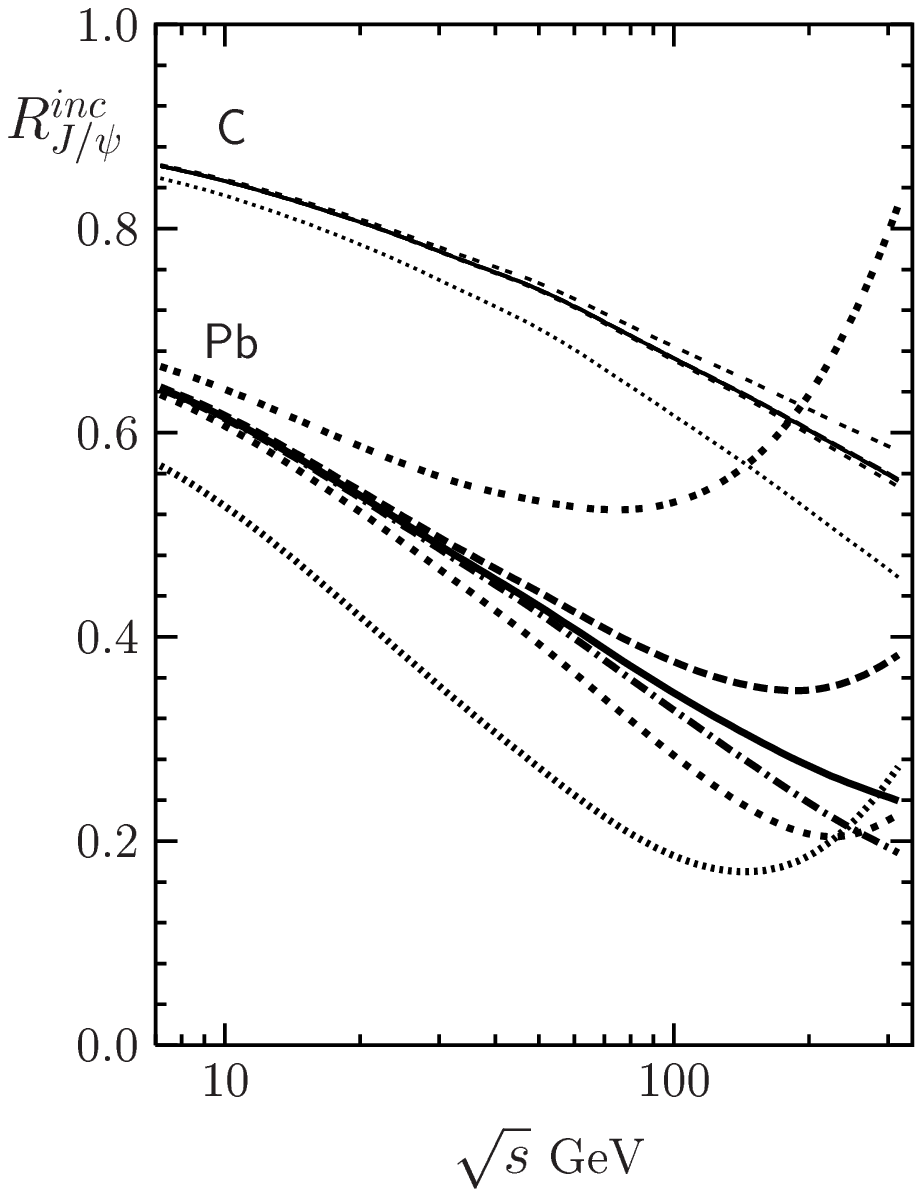}\hskip5mm
  \PSfig{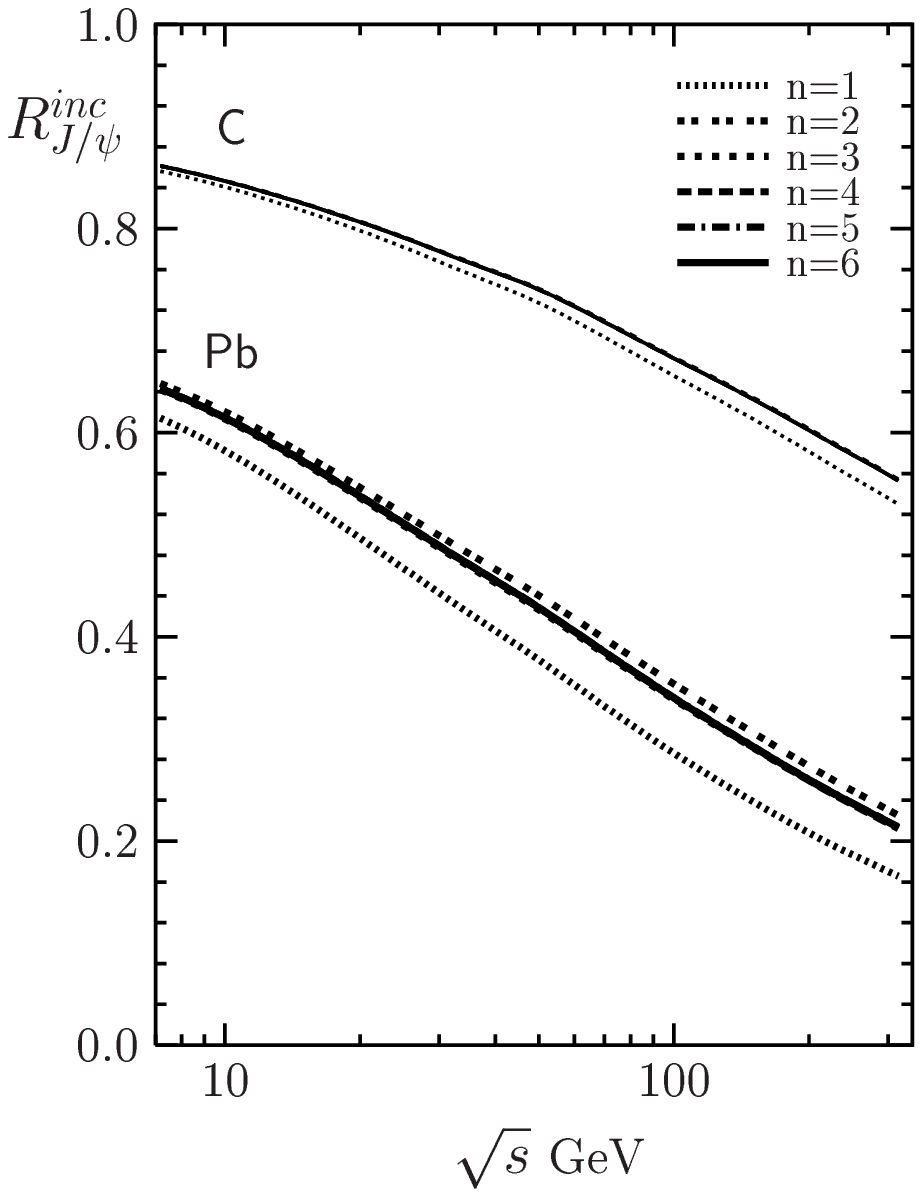}
}
\Caption{
  \label{FigResum}
  Convergence of the series \Ref{ResumT} (left) and resummation
  in the exponent \Ref{ResumE} (right) for incoherent $\Jpsi$
  production on nuclei ($Q^2=0$, gluon shadowing and $l_c$
  corrections are not included) for different orders of
  summation on $f_n$.
}
\EF
\FloatBarrier

% -----------------------------------------------------------

\end{document}